\documentclass[journal]{IEEEtran}

\usepackage{amsmath,amsthm,amsfonts,amssymb,times,bbm,graphicx}

\newtheorem{theorem}{Theorem}
\newtheorem{proposition}[theorem]{Proposition}

\newtheorem{lemma}[theorem]{Lemma}
\newtheorem{corollary}[theorem]{Corollary}
\newtheorem{definition}[theorem]{Definition}

\def\RR{\mathbbm{R}}
\def\CC{\mathbbm{C}}
\def\FF{\mathbbm{F}}

\def\EE{\mathbbm{E}}

\def\P{\mathcal{P}}
\def\R{\mathcal{R}}

\def\Id{\mathbbm{1}}
\def\id{\mathrm{id}}

\def\Xb{\mathbf{X}}

\DeclareMathOperator{\tr}{tr}

\DeclareMathOperator{\rank}{rank}
\DeclareMathOperator{\range}{range}

\DeclareMathOperator{\sign}{sgn}

\begin{document}

\title{Recovering Low-Rank Matrices From Few Coefficients In Any
Basis}

\author{David Gross
\thanks{This work was supported by the EU (CORNER).}%
\thanks{D.~Gross is with the Institute for Theoretical Physics,
Leibniz University Hannover, 30167 Hannover, Germany and with the Institute of
Theoretical Physics, ETH Zurich, 8093 Zurich, Switzerland. e-mail: see
www.phys.ethz.ch/\~{ }dagross}}

\maketitle

\begin{abstract} 
	We present novel techniques for analyzing the problem of low-rank
	matrix recovery. The methods are both considerably simpler and more
	general than previous approaches. It is shown that an unknown
	$n\times n$ matrix of rank $r$ can be efficiently reconstructed from
	only $O(n r \nu\,\ln^2n)$ randomly sampled expansion coefficients
	with respect to any given matrix basis. The number $\nu$ quantifies
	the ``degree of incoherence'' between the 	unknown matrix and the
	basis.  Existing work concentrated mostly on the problem of ``matrix
	completion'' where one aims to recover a low-rank matrix from
	randomly selected matrix elements.  Our result covers this situation
	as a special case. The proof consists of a series of relatively
	elementary steps, which stands in contrast to the highly involved
	methods previously employed to obtain comparable results. In cases
	where bounds had been known before, our estimates are slightly
	tighter.  We discuss operator bases which are incoherent to all
	low-rank matrices simultaneously. For these bases, we show that $O(n
	r \nu \ln n)$ randomly sampled expansion coefficients suffice to
	recover any low-rank matrix with high probability. The latter bound
	is tight up to multiplicative constants.  
\end{abstract}


\section{Introduction}

We consider the problem of efficiently recovering a low-rank matrix
from a small number of expansion coefficients with respect to some
basis in the space of matrices. Related questions have recently
enjoyed a substantial amount of attention (c.f.\
\cite{
recht_guaranteed_2007,
candes_exact_2009,candes_power_2009,
candes_matrix_2009,
singer_uniqueness_2009,
keshavan_matrix_2009,
wright_robust_2009} for a highly
incomplete list of references).

To get some intuition for the problem, note that one needs roughly
$rn$ parameters to specify an  $n\times n$-matrix $\rho$ of rank $r$.
Therefore, it might be surmised that about the same number of
expansion coefficients of $\rho$ (with respect to some fixed matrix
basis) are sufficient to uniquely specify $\rho$ within the set of
low-rank matrices. It is by far less clear whether $\rho$ can be
recovered from this limited set of coefficients in a computationally
tractable way.

Low-rank matrix recovery may be compared to a technique studied under
the name of \emph{compressed sensing} \cite{donoho_compressed_2006,
candes_robust_2006, candes_near-optimal_2006}. In its simplest
version, the task there is to recover a sparse vector from few Fourier
coefficients. Informally, the property of having a low rank is the
``non-commutative analogue'' of sparsity. In this sense, one may think
of the matrix recovery problem as a non-commutative version of
compressed sensing.

This field of research was started in earnest with the results in
\cite{candes_exact_2009,candes_power_2009}.  There, it was shown that
surprisingly, reconstructing a rank-$r$ matrix from only $O(n r
\operatorname{polylog(n)})$ randomly selected matrix elements can be
done efficiently employing a simple convex optimization algorithm.
These findings were partly inspired by methods used earlier in
compressed sensing \cite{candes_robust_2006,candes_near-optimal_2006}.

The results presented in \cite{candes_exact_2009,candes_power_2009}
were as spectacular as they were difficult to prove; the tighter
bounds in \cite{candes_power_2009} required dozens of pages. At the
same time, the proof techniques seemed to be tailored to the fact that
matrix elements, as opposed to more general expansion coefficients,
had been sampled. 

In \cite{gross_quantum_2009} the present author and collaborators
developed new methods for analyzing low-rank matrix recovery problems.
The work was motivated by the desire to prove analogues of
\cite{candes_exact_2009,candes_power_2009} applicable to certain
problems in quantum mechanics. Three main improvements were achieved.
Most importantly, the mathematical effort for obtaining near-optimal
bounds on the number of coefficients needed to determine a low-rank
matrix was cut dramatically, with a condensed (but complete) version
of the proof fitting on a single page. Also, the new arguments depend
much less on the specific properties of the basis used.  Lastly, in
some situations, the bounds obtained are tighter than those presented
previously. In some cases, the gap between lower and upper bounds is
reduced to a multiplicative constant.

The present paper builds on the methods of \cite{gross_quantum_2009}.
It aims to make them accessible to readers not accustomed to the
language of quantum information theory, supplies many details missing
in \cite{gross_quantum_2009} due to space limitations, generalizes
the results to arbitrary operator bases, and provides tighter
estimates.

\subsection{Setting and main results}
\label{sec:setting}

Throughout the main part of this paper the word ``matrix'' will be
used to mean ``\emph{Hermitian} matrix'' (or, equivalently,
``symmetric matrix'', if one prefers to work over the real numbers).
Our methods work more naturally in this setting, and a lack of
Hermiticity would just be a technical problem obscuring the essence of
the argument. In fact little generality is lost. In
Section~\ref{sec:nonHermitian}, we describe a straight-forward way for
translating any non-Hermitian matrix recovery problem to a Hermitian
one.  Therefore, in essence, all our results include this more general
case.

The unknown rank-$r$ matrix to be recovered will be denoted by $\rho$.
On the space of Hermitian matrices, we use the Hilbert-Schmidt inner
product
$(\sigma_1, \sigma_2)=\tr(\sigma_1^\dagger\sigma_2)$. We assume that some
ortho-normal basis $\{w_a\}_{a=1}^{n^2}$ with respect to this inner
product has been chosen (referred to as an \emph{operator basis}).
Thus, $\rho$ can be expanded as
\begin{equation*}
	\rho = \sum_{a=1}^{n^2} (w_a, \rho)\, w_a.
\end{equation*}

The question addressed below is: \emph{given that $\rank\rho\leq r$, how
many randomly chosen coefficients $(w_a, \rho)$ do we need to know,
before we can efficiently reconstruct $\rho$?}

In order to perform the reconstruction, we will utilize the algorithm
employed in
\cite{candes_robust_2006,recht_guaranteed_2007,candes_exact_2009,candes_power_2009}.
Let $\Omega\subset[1,n^2]$ be a random set of size $m$. Assume that we
know
the coefficients $(w_a, \rho)$ for all $a\in \Omega$. The algorithm
simply consists of performing the following (efficiently implementable)
convex optimization over the space of matrices:
\begin{eqnarray}
	\min && \|\sigma\|_1 \label{eqn:optimization} \\
	\text{subject to} && (\sigma,w_a) = (\rho,w_a), \quad \forall\,a \in \Omega.
	\nonumber
\end{eqnarray}
Above, $\|\sigma\|_1$ is the \emph{trace-norm}
(also \emph{Schatten 1-norm} or \emph{nuclear norm}), i.e.\ 
the sum of the singular values of $\sigma$. Let $\sigma^\star$ be a
solution of the optimization. Theorem~\ref{thm:main}
quantifies the probability (with respect to the sampling process) of
$\sigma^\star$ being unique and equal to $\rho$, as a function of the
the number $m$ of coefficients revealed. 

It is clear that the algorithm will perform poorly if $\rho$ has very
few non-zero expansion coefficients with respect to the basis
$\{w_a\}$ \cite{candes_exact_2009}.  To avoid such a situation, we
must ensure that a typical coefficient will contain ``enough
non-trivial information'' about $\rho$. That is the content of the
various notions of ``incoherence'' which have been proposed
\cite{candes_exact_2009,candes_power_2009}.  Our definition of
incoherence is stated below. It is closely related to, but more
general than, the parameter $\mu$ used in
\cite{candes_exact_2009,candes_power_2009}. In particular, going
beyond previously published situations, we find that there are certain
bases with the property that \emph{any} low-rank matrix is incoherent
with respect to them.

To state the results more precisely, we need to introduce some
notation. (We try to follow \cite{candes_exact_2009} as closely as
possible).
Let $U=\range\rho$ be the row space of $\rho$ (which is equal to its
column space, due to Hermiticity). Let $P_U$ be the orthogonal
projection onto $U$. The space of matrices 
\begin{equation}\label{eqn:T}
	T=\{ \sigma \,|\, (\Id-P_U) \sigma(\Id-P_U) = 0 \}
\end{equation}
whose compression to $\ker\rho$ vanishes
will play an important role ($\Id$ is the identity matrix; see also
Fig.~\ref{fig:Delta}). The map
\begin{equation*} 
	\P_T: \sigma\mapsto P_U \sigma+ \sigma P_U - P_U \sigma P_U.
\end{equation*} 
projects\footnote{
	We will use calligraphic $\P$'s for matrix-valued projections, and
	roman $P$'s for vector-valued projections.
} 
onto $T$.  Whenever there is little danger of confusion, we will not
make the dependency of $T, \P_T$ and other objects on $\rho$ explicit
in our notation.

Recall the definition of the sign function: $\sign(x) = x/|x|$ for $x
\neq 0$ and $\sign(0)=0$. Below, we will apply the sign function (and
other real functions) to Hermitian matrices. Expressions like
$\sign\sigma$ are to be understood in terms of the usual ``functional
calculus''. I.e.\ $\sign\sigma$ is the matrix which is diagonal in the
same basis as $\sigma$, but with eigenvalues $\sign(\lambda_i)$, where
the $\lambda_i$ are the eigenvalues of $\sigma$.

The unadorned norm $\|\sigma\|$ of a matrix $\sigma$ refers to the
\emph{operator norm} (or \emph{spectral norm}): 
the largest singular value. The 2-norm (also \emph{Frobenius norm}) is
$\|\sigma\|_2=\tr(\sigma\sigma)^{1/2}$.

We can now state our definition of coherence.

\begin{definition}[Coherence]\label{def:coherence}
	The $n\times n$-matrix $\rho$ has \emph{coherence} $\nu$ with
	respect to an operator basis $\{w_a\}_{a=1}^{n^2}$ if either
	\begin{eqnarray}\label{eqn:coherenceNorm}
		\max_a \|w_a\|^2 &\leq& \nu \frac {1}{n}
	\end{eqnarray}
	or the two estimates
	\begin{eqnarray}\label{eqn:coherence1}
		\max_a \|\P_T w_a\|_2^2 &\leq& 2 \nu \frac{r}{n},  \\
		\max_a \,(w_a, \sign\rho)^2 &\leq& \nu \frac{r}{n^2}
		\label{eqn:coherence2}
	\end{eqnarray}
	hold.
\end{definition}

Let $\{e_1, \dots, e_n\}$ be the standard basis in $\CC^n$. The
(non-Hermitian) \emph{standard operator basis} is
$\{e_i e_j^\dagger\}_{i,j=1}^n$, where $e_i
e_j^\dagger$ is the matrix whose only non-zero element is a $1$ at the
intersection of the $i$th row and the $j$th column.  The best
previously known result seems to be this:

\begin{theorem}[{\cite[Thm.~1.1]{candes_power_2009}}]
	Let $\rho$ be a rank-$r$ matrix with coherence $\nu$ with respect to
	the standard operator basis.  Let $\Omega\subset[1,n^2]$ be a random
	set of size $|\Omega|\geq O(n r \nu^4 \ln^2n)$. Then the solution
	$\sigma^\star$ of the optimization problem (\ref{eqn:optimization})
	is unique and equal to $\rho$ with probability at least $1-n^{-3}$.
\end{theorem}

Our main theorem works for arbitrary operator bases, improves the
$\nu$-dependency and will turn out to be easier to prove.

\begin{theorem}[Main result]\label{thm:main}
	Let $\rho$ be a rank-$r$ matrix with coherence $\nu$ with respect
	to an operator basis $\{w_a\}_{a=1}^{n^2}$.
	Let $\Omega\subset[1,n^2]$ be a random set of size $|\Omega|\geq O(n r \nu
	(1+\beta) \ln^2n)$.
	Then the solution $\sigma^\star$ of the
	optimization problem (\ref{eqn:optimization}) is unique and equal to
	$\rho$ with probability at least $1-n^{-\beta}$.
\end{theorem}

The precise condition on $|\Omega|$ for the statement in
Theorem~\ref{thm:main} to hold is
\begin{equation*}
	|\Omega|> \log_2(2n^2\sqrt r)
	64 \nu(\ln(4n^2)+\ln(9\log_2n) +
	\beta \ln n)rn.
\end{equation*}
(No attempt has been made to optimize the constants appearing in this
expression.) 
In the expositional part of this paper, we will frequently employ the
``big-Oh''-notation\footnote{
	We write $|\Omega|\geq O(f(n,r,\nu,\beta))$
	if there is a constant $C$ such that for $n$ large enough and for all
	$\nu,\beta$ and $r\leq n$, it holds that $|\Omega|\geq C f(n,r,\nu,\beta)$.
}
to give simplified accounts of otherwise complex expressions. However,
in the more technical sections, it will be shown that all statements
hold for any finite $n$ (and not just asymptotically, as the
$O$-notation might suggest) and all constants will be worked out
explicitly.

We remark that the only property of the basis $\{w_a\}$ itself that
has entered the discussion so far is its operator norm $\max_a
\|w_a\|$. Intuitively, the reason is easily understood: matrices with
small operator norm are ``incoherent'' to all low-rank matrices
simultaneously. More precisely: if $\rho$ is a matrix of rank $r$,
normalized such that $\|\rho\|_2 = 1$, then
H\"older's inequality
for matrices \cite[Corollary IV.2.6]{bhatia_matrix_1997}
gives the estimate
\begin{equation}\label{eqn:holder}
	|(w, \rho)|^2 
	\leq \|w\|^2\,\|\rho\|_1^2 
	\leq \|w\|^2\,r
\end{equation}
for any matrix $w$.  Hence the squared overlap on the left hand side
is small if both $r$ and $\|w\|$ are.
As a corollary, we can actually derive (\ref{eqn:coherence1}) from
(\ref{eqn:coherenceNorm}). Indeed
\begin{eqnarray*}
	\|\P_T w_a\|_2^2 
	&=& \sup_{\sigma\in T, \|\sigma\|_2=1} (w_a, \sigma)^2
	\leq \|w_a\|^2 \|\sigma\|_1^2 \nonumber \\
	&\leq& \|w_a\|^2\, 2r \|\sigma\|_2^2
	\leq 2 \nu \frac rn
\end{eqnarray*}
(having used the simple fact that $\max_{\sigma\in
T}(\rank\sigma)=2r$).

Equation (\ref{eqn:holder}) has a well-known analogue in compressed
sensing
\cite{donoho_compressed_2006,candes_robust_2006,candes_near-optimal_2006}.
There, one uses the fact that ``vectors with small entries'' are
incoherent to ``sparse vectors''. Indeed, if $\sigma_1, \sigma_2$ are
vectors, $\|\sigma_1\|$ is taken to be the supremum norm (i.e. the
absolute value of the largest component of $\sigma_1$) and
$\rank\sigma_2$ is the number of non-zero entries of $\sigma_2$, then
Eq.~(\ref{eqn:holder}) remains true. The best-known example of a basis
consisting of vectors with small supremum norm is the Fourier basis.
Motivated by this analogy, we will refer to operator bases 
fulfilling (\ref{eqn:coherenceNorm}) as
\emph{Fourier-type bases}. Arguably, from a mathematical point of
view, they form the most natural setting for low-rank matrix
recovery\footnote{
To the best knowledge of the author, the first researcher who clearly
appreciated the significance of the basis' operator norm was Y.-K.\
Liu. He proved that some of the bounds in \cite{candes_exact_2009} continue
to hold for all low-rank matrices, if -- instead of matrix elements -- one
samples expansion coefficients with respect to a certain unitary
operator basis \cite{liu__2009}.
}.

We will prove Theorem~\ref{thm:main} for Fourier-type bases first and then present
two relatively simple modifications which allow us to cover the
general case. 

In later sections we will refine the analysis for Fourier-type bases,
arriving at Theorem~\ref{thm:tight}. Asymptotically, the estimate is
tight up to multiplicative constants.

\begin{theorem}[Tighter bounds for Fourier-type bases]\label{thm:tight}
	Let $\rho$ be a rank-$r$ matrix and suppose that
	$\{w_a\}$ is an operator basis fulfilling $\max_a \|w_a\|^2\leq
	\frac{\nu}{n}$.
	Let $\Omega\subset[1,n^2]$ be a random set.
	Then the solution $\sigma^\star$ to the
	optimization problem (\ref{eqn:optimization}) is unique and equal to
	$\rho$ with probability of failure smaller than
	$e^{-\beta}$,
	provided that 
	\begin{equation*}
		|\Omega|\geq O(n r \nu (\beta+1)\ln n).
	\end{equation*}
\end{theorem}

Comparable bounds were known before in situations where the operator
basis itself was drawn randomly (as opposed to a random subset from
any given basis) \cite{recht_guaranteed_2007} or under additional
assumptions on the spectrum of $\rho$ \cite{keshavan_matrix_2009}.
However, this seems to be the first time the optimal $\log$-factor in
the bound on $|\Omega|$ has been proven to be achievable in a matrix
recovery problem, where the involved basis and unknown matrix were
neither randomized nor subject to constraints beyond their rank.

\subsection{Examples}
\label{sec:examples}

\subsubsection{Matrix completion}

We apply Theorem~\ref{thm:main} to the special case of matrix completion, as
treated in 
\cite{candes_exact_2009,
candes_power_2009,
keshavan_matrix_2009}. Denote the standard basis in
$\CC^n$ by $\{e_i\}_{i=1}^n$ and let $\{e_i e_j^\dagger\}_{i,j}^{n}$
be the standard operator basis.
Set $U=\range\rho$ and let $P_U$ be the orthogonal projection
onto $U$. Assume that $\rho$ fulfills
\begin{equation*}
	\max_i \|P_U e_i\|_2^2 \leq \mu_1 \frac{r}{n},
	\qquad
	\max_{i,j}\, |\langle e_i, \sign\rho\, e_j\rangle| \leq 
	\mu_2 \sqrt{\frac{r}{n^2}}.
\end{equation*}
(angle brackets refer to the standard inner product in $\CC^n$).

Because we work in the setting of Hermitian matrices, it holds that
\begin{equation*}
	\langle e_i, \rho\,e_j\rangle
	=
	\langle e_j, \rho\,e_i\rangle^*,
\end{equation*}
so that every time one matrix element is revealed, we additionally
obtain knowledge of the transposed one. Accordingly, the Hermitian
analogue of sampling matrix elements is sampling expansion
coefficients with respect to the basis $\{w_a\}$ of matrices of the
form
\begin{equation}\label{eqn:matrixelements}
	1/\sqrt 2\big( e_i e_j^\dagger + e_j e_i^\dagger),
	\qquad
	{\rm i}/\sqrt 2\big( e_i e_j^\dagger - e_j e_i^\dagger)
\end{equation}
for $i<j$, together with the matrices $e_i e_i^\dagger$ supported on the
main diagonal.

One now simply verifies
\begin{equation*}
	\max_a \|\P_T w_a\|_2^2 \leq 2 \mu_1\frac{r}n,
	\qquad
	\max_a \,(w_a, \sign\rho)^2 \leq2 \mu_2^2 \frac{r}{n^2}.
\end{equation*}
Thus, Theorem~\ref{thm:main} is applicable with $\nu=\max\{\mu_1, 2 \mu_2^2\}$.

\subsubsection{Unitary operator bases}
\label{sec:unitaryBases}

We briefly comment on bases with minimal operator norm. Let $\{w_a\}$
be an ortho-normal basis in the space of matrices. At this point, we
do not assume that the basis is Hermitian. Denote the singular values
of $w_a$ by $s_i(w_a)$. Since
\begin{equation*}
	1=\|w_a\|_2^2=\sum_{i=1}^n (s_i(w_a))^2,
\end{equation*}
it follows that $\|w_a\|^2=\max_i s_i^2(w_a)\geq \frac1n$. Therefore
$\nu=1$ is the best possible value in (\ref{eqn:coherenceNorm}). It
is achieved exactly if $\sqrt n \,w_a$ is unitary for every
$a\in[1,n^2]$. Such \emph{unitary operator bases} have been studied in
some detail (see e.g.~\cite{werner_all_2001}).

A standard example with manifold applications is the
\emph{Pauli (operator) basis}. 
For $n=2$ it is given by $w_a = \frac1{\sqrt 2} \sigma_a$, where 
\begin{eqnarray*}
	\sigma_1=
	\left(
		\begin{array}{cc}
			0 & 1 \\
			1 & 0
		\end{array}
	\right),
	&&\qquad 
	\sigma_2=
	\left(
		\begin{array}{cc}
			0 & -{\rm i} \\
			{\rm i} & 0
		\end{array}
	\right), \\
	\sigma_3=
	\left(
		\begin{array}{cc}
			1 & 0 \\
			0 & -1
		\end{array}
	\right),
	&&\qquad 
	\sigma_4=
	\left(
		\begin{array}{cc}
			1 & 0 \\
			0 & 1
		\end{array}
	\right)
\end{eqnarray*}
are the \emph{Pauli matrices}. The $\sigma_i$'s have eigenvalues
$\{\pm 1\}$ and are thus both unitary and Hermitian. The Pauli basis
$\{w_a^{(k)}\}$
for matrices acting on $(\CC^2)^{\otimes k} \simeq \CC^{2^k}$ is
defined as the $k$-fold tensor product basis with factors the
$\{w_a^{(1)}\}$ above.

The bases $\{w_a^{(k)}\}$ possess an exceedingly rich structure which
is at the heart of many central results in quantum information theory
(see e.g.\
\cite{nielsen_quantum_2000,gottesman_stabilizer_1997,montanaro_quantum_2008};
for a brief introduction see \cite{gross_hudsons_2006}). We will make
use of the existing theory to prove lower bounds on $|\Omega|$ in
Section~\ref{sec:lower}.  

The Pauli basis is a commonly used ingredient in experimental
quantum-state tomography---a fact which initially motivated this work.

\subsection{Intuition}
\label{sec:intuition}

\begin{figure}
	\begin{center}
		(a)
		\includegraphics{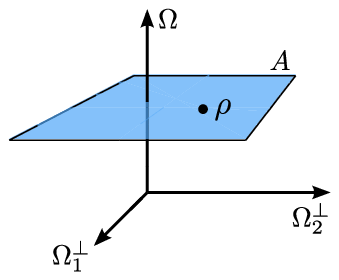}
		\vspace{.5cm}
		(b)
		\includegraphics{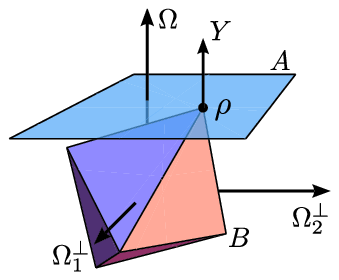}
	\end{center}
	\caption{\label{fig:affine} 
		(a) The unknown matrix $\rho$ is an element of a $n^2$-dimensional
		linear space. The axis labeled $\Omega$ represents all the
		coordinates of $\rho$ known to us. We have no information about
		the projection of $\rho$ onto the orthogonal directions,
		represented by the axes labeled $\Omega^\bot$. Thus the set of
		matrices compatible with the coefficients known to us forms an
		affine space $A$, parts of which are indicated in the figure.  ---
		(b) The convex program (\ref{eqn:optimization}) recovers $\rho$ if
		it is the unique minimizer of the trace-norm
		restricted to $A$. This is certainly the case if $A$ is contained
		in a supporting hyperplane at $\rho$ of the trace-norm ball
		$B=\{\sigma \,|\, \|\sigma\|_1 \leq \|\rho\|_1\}$. In other
		words, there must be a normal vector $Y$ to a supporting
		hyperplane of $B$ at $\rho$, such that $Y$ is also normal to $A$.
		In the language of convex optimization, $Y$ is referred to as a
		\emph{dual certificate}.
	}
\end{figure}

The basic intuition underlying our results differs little from previous
approaches
\cite{candes_exact_2009,candes_power_2009,recht_guaranteed_2007,
candes_matrix_2009}. For the sake of
being self-contained, we still give a brief non-technical account of
some aspects we find essential.  (Technical differences to existing
publications are outlined in the next section.)

Consider the sketch in Fig.~\ref{fig:affine}(a) (partly inspired by
\cite{candes_matrix_2009}). The matrix $\rho$ is an element of an
$n^2$-dimensional linear space. The axis labeled $\Omega$ in the diagram
represents the roughly $O(rn)$ coordinates we have information about, i.e.\
the space spanned by the $\{w_a\,|\, a \in \Omega\}$. As the $n^2-O(r
n)$ remaining coordinates (denoted by $\Omega^\bot$) are unknown,
there is a large affine space of matrices compatible with the
available information. We have to specify an algorithm which picks one
point from this high-dimensional affine space, and prove that our
choice is identical to $\rho$ with high probability.

Since we are looking for a low-rank object, it would be natural to
choose the lowest-rank matrix in the affine space of all matrices
compatible with the information we have. However, minimizing the rank
over an affine space is in general NP-hard
\cite{natarajan_sparse_1995}.  To get around this problem, we employ
the \emph{trace heuristic}, which stipulates that minimizing the
trace-norm is a good proxy for rank minimization (see e.g.\
\cite{mesbahi_rank_1997, fazel_rank_2001}). The resulting optimization
problem (\ref{eqn:optimization}) is an efficiently solvable
semi-definite program. 

The objective thus becomes proving that the trace-norm restricted to
the affine plane has a strict and global minimum at $\rho$ (Fig.\
\ref{fig:affine}(b)). Thus, if $\rho+\Delta\neq \rho$ is any matrix in the
affine plane, we need to show that 
\begin{equation}\label{eqn:minimum}
	\|\rho+\Delta\|_1>\|\rho\|_1.
\end{equation}
A short handwaving argument indicates that adding a generic deviation
$\Delta$ to a low-rank $\rho$ is indeed likely to increase the
trace-norm.

To see why, recall that the trace-norm of a matrix is larger than the
sum of the absolute values of the elements on the main diagonal
\cite{marshall_inequalities_1979}. We will apply this estimate to $\rho+\Delta$
expressed in some eigenbasis of $\rho$. Let $\rho_1, \dots, \rho_r$ be
the eigenvalues of $\rho$. Then
\begin{eqnarray}
	\|\rho+\Delta\|_1 
	&\geq& \sum_{i=1}^r |\rho_i+\Delta_{i,i}| + \sum_{i=r+1}^n |\Delta_{i,i}|
	\nonumber \\
	&\geq& \|\rho\|_1 + \sum_{i=1}^r (\sign r_i) \Delta_{i,i} + \sum_{i=r+1}^n
	|\Delta_{i,i}|. \label{eqn:majorized}
\end{eqnarray}
For \emph{generic} deviations $\Delta$, we expect that the
$\Delta_{i,i}$ all have comparable magnitudes. Therefore, as long as
$r\ll n$, the second sum in (\ref{eqn:majorized}) will dominate the
first one as required.

The ``only'' difficulty faced in this paper consists in proving that
$\|\rho+\Delta\|_1>\|\rho\|_1$ holds not just for generic matrices
$\rho+\Delta$ in the aforementioned affine plane, but for all such
elements simultaneously. Key to that will be a simple concept from
convex optimization theory: a \emph{dual certificate}
\cite{bertsekas_convex_2003,
candes_robust_2006,
candes_exact_2009,
candes_power_2009}.
By that we mean a matrix $Y$ such that
\begin{equation}\label{eqn:dual}
	\|\rho+\Delta\|_1 > \|\rho\|_1 + (Y, \Delta)
\end{equation}
for $\Delta\neq 0$.
If we can find such a $Y$ which is also normal to the affine plane
(c.f.\ Fig.~\ref{fig:affine}(b)), then the inner product above
vanishes and (\ref{eqn:dual}) implies (\ref{eqn:minimum}). 

The main contribution of this work is an improved and generalized
construction of an (approximate) dual certificate $Y$.

\subsection{Novel approaches}
\label{sec:novel}

For readers well-accustomed to previous work, we shortly list some
main technical differences.
\begin{enumerate}
	\item 
	We employ an i.i.d.\ sampling process (sampling with replacement) to
	chose the revealed coefficients. This contrasts with the
	``Bernoulli'' scheme used before \cite{candes_exact_2009,candes_power_2009}. 	

	\item
	At two different points in the proof (Section~\ref{sec:scales},
	Section~\ref{sec:smallBases}), we make use of a powerful
	large-deviation estimate for matrix-valued observables. This
	(so far under-appreciated?) \emph{operator Chernoff bound} has been
	proven in \cite{ahlswede_strong_2002}. 

	\item
	In the language of \cite{candes_exact_2009}, when constructing a ``dual
	certificate''-type matrix $Y$ we note that it is sufficient to
	demand $\|\P_T E - Y\|_2$ be small, as opposed to zero
	(Section~\ref{sec:linearize}). The former
	is simpler to ascertain than the latter. 

	\item
	We construct a particular matrix-valued random process
	(descriptively
	called the ``golfing scheme''), which converges to the certificate
	$Y$ exponentially fast (Section~\ref{sec:smallBases}).
\end{enumerate}

\subsection{Previous versions of this result and some related work}
\label{sec:previous}

This work grew out of an effort to translate the results of
\cite{candes_exact_2009, candes_power_2009} to the problem of
quantum-state tomography, where bases of Fourier-type matrices
naturally occur.  The project turned out to lead to more general
results than anticipated, producing the methods presented in this
paper.

We first published these results in \cite{gross_quantum_2009}, a short
paper written with a physics audience in mind. This pre-print contains
all the main ideas of the current work, and a complete proof of
Theorem~\ref{thm:main} for Fourier-type bases (the case of interest in quantum
tomography). We announced in \cite{gross_quantum_2009} that a more
detailed exposition of the new method, applying to the general
low-rank matrix recovery problem with respect to arbitrary bases, was
in preparation.

Before this extended version of \cite{gross_quantum_2009} had been
completed, another pre-print \cite{recht_simpler_2009} building on
\cite{gross_quantum_2009} appeared. The author of
\cite{recht_simpler_2009} presents our methods in a language more
suitable for an audience from mathematics or information theory. He
also presents another special case of the results announced in
\cite{gross_quantum_2009}: the reconstruction of low-rank matrices
from randomly sampled matrix elements. The main proof techniques in
\cite{recht_simpler_2009} are identical to those of
\cite{gross_quantum_2009}, with two exceptions. First, the author
independently found the same modification we are using here to extend
the methods from Fourier-type matrices to bases with larger operator
norm (his Lemma~3.6, our Lemma~\ref{lem:stayIncoherent}). Second, his
proof works more directly with non-Hermitian matrices, and gives
tighter bounds in the case of non-square matrices.

A more detailed version of \cite{gross_quantum_2009} focusing on
physics issues will appear elsewhere \cite{becker__2009}.

\section{Main proof}

\subsection{The ensemble}
\label{sec:ensemble}

Let $A_1, \dots, A_m$ be random variables taking values in $[1,n^2]$.
Their distribution will be specified momentarily. Important objects in
our analysis are the matrix-valued random variables $w_{A_i}$. 
The \emph{sampling operator} is
\begin{equation}\label{eqn:r}
	\R: \sigma \mapsto \frac{n^2}{m} \sum_{i=1}^m w_{A_i}\,
	(w_{A_i},\sigma).
\end{equation}
Below, we will analyze the semi-definite program
\begin{eqnarray}
	\min && \|\sigma\|_1 \label{eqn:optimizationR} \\
	\text{subject to} && \R \sigma = \R \rho.  \nonumber
\end{eqnarray}

If the $A_i$'s correspond to $m$ samples drawn from $[1,n^2]$
\emph{without} replacement, the programs (\ref{eqn:optimization}) and
(\ref{eqn:optimizationR}) are equivalent. One can also consider the
situation where the $A_i$'s are i.i.d.\ random variables, describing
sampling \emph{with} replacement. Due to independence, the latter
situation is much easier to analyze. Independence also implies the
possibility of \emph{collisions}\footnote{
	By the ``birthday paradox'', such collisions are very likely to
	occur.
} (i.e.\ $A_i=A_j$, for $i\neq j$).
In the presence of collisions, fewer than $m$ distinct coefficients
will contribute to (\ref{eqn:optimizationR}). It is thus plausible
(and will be confirmed below) that any upper bound on the probability
of failure of the i.i.d.\ scheme is also valid for
(\ref{eqn:optimization}). From now on, we will therefore assume that
the $A_i$'s are independent and uniformly distributed.

To state the obvious: the solution $\sigma^\star$ to
(\ref{eqn:optimizationR}) is unique and equal to $\rho$ if and only if
any non-zero deviation $\Delta=\sigma-\rho$ from $\rho$ is either
\emph{infeasible}
\begin{equation}\label{eqn:infeasible}
	\R\Delta \neq 0,
\end{equation}
or causes the trace-norm to increase
\begin{equation}\label{eqn:tracenorm}
	\|\rho+\Delta\|_1 > \|\rho\|_1.
\end{equation}
The two conditions (\ref{eqn:infeasible}), (\ref{eqn:tracenorm}) have
a very different mathematical flavor. Section~\ref{sec:scales}
concentrates on the first one, while the second one is more central in
the remainder.

Using (\ref{eqn:infeasible}), one can give a simple proof of our
earlier remark that sampling \emph{with} replacement can only decrease
the probability of recovering $\rho$:
\begin{IEEEproof}
	Let $p_{\mathrm{with}}(m), p_{\mathrm{wout}}(m)$ be the
	probabilities that the solution of (\ref{eqn:optimizationR}) equals
	$\rho$, if the $A_1, \dots, A_m$ are sampled, respectively, with or
	without replacement.

	Let $\R'$ be defined as in (\ref{eqn:r}), but with the sum extending
	only over distinct samples $A_i\neq A_j$ (denote the number of
	distinct samples by $m'$). Then $\ker\R'=\ker\R$, and consequently
	(\ref{eqn:infeasible}) is true for $\R$ iff it is true
	for $\R'$.

	Thus, the probability that the solution to (\ref{eqn:optimizationR})
	equals $\rho$ is the same as the probability that the
	solution of 
	\begin{equation}\label{eqn:optmizationR'}
		\min \|\sigma\|_1 \qquad \textrm{subject to}\quad \R'\sigma=\R'\rho
	\end{equation}
	equals $\rho$. But, conditioned on any value of $m'$, the
	distribution of $\R'$ is the same as the distribution of a sampling
	operator drawing $m'$ basis elements without replacement. Hence
	\begin{equation*}
		p_{\mathrm{with}}(m) 
		=\EE_{m'}[p_{\mathrm{wout}}(m')]
		\leq p_{\mathrm{wout}}(m),
	\end{equation*}
	since $m'\leq m$ and clearly 
	$p_{\mathrm{wout}}(m') \leq p_{\mathrm{wout}}(m)$ 
\end{IEEEproof}

The i.i.d.\ scheme used in the present papers contrasts with the
``Bernoulli model'' employed in previous works
\cite{candes_near-optimal_2006,candes_exact_2009,candes_power_2009}.
There, every number $a\in[1,n^2]$ is included in $\Omega$ with
probability $m/n^2$. The slight advantage of our approach is that the
random variables $(w_{A_i}, \rho)$ are identically distributed, in
addition to being independent. Also, the random process analyzed here
never obtains knowledge of \emph{more} than $m$ coefficients, while
this does happen in the Bernoulli model with finite probability. On
the downside, the possibility of incurring collisions has some
technical drawbacks, e.g.\ it means that $\R$ will in general not be
proportional to a projection. 

\emph{Note added:} after the pre-print version of this paper had been
submitted, V.~Nesme and the author noted that existing arguments
pertaining to sampling without replacing of real-valued random
variables \cite[Chapter 12]{marshall_inequalities_1979} remain valid in the
non-commutative case \cite{gross__2009}. In particular, all large
deviation bounds derived below under the assumption of independently
chosen coefficients continue to hold for $A_i$'s sampled without
replacement. While we will not make use of these observations in the
present paper, we note that they can be used to slightly improve the
bounds given below. Details are in \cite{gross__2009}.

\subsection{Further layout of proof and notation}
\label{sec:notation}

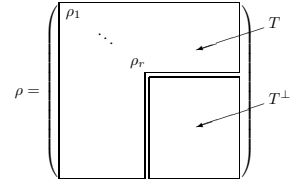
\begin{figure}
	\begin{center}
		\setlength{\unitlength}{1mm}
		\scalebox{.6}{
			\begin{picture}(60,50)
				\put(-6.5,20){
					$\rho=\left(
						\begin{array}{llll}
							\rho_1 			& 					&					& \hspace{1.6cm} \\
													&		\ddots 	&					& \\ 
													&						& \rho_r 	& \\
							\vspace{2cm}&						&					& 
						\end{array}
					\right)$
				}
				\put(4.5,1.5){\line(0,1){39.0}}
				\put(4.5,40.5){\line(1,0){40}}
				\put(44.5,40.5){\line(0,-1){15.5}}
				\put(44.5,25.0){\line(-1,0){21}}
				\put(23.5,25.0){\line(0,-1){23.5}}
				\put(4.5,1.5){\line(1,0){19.0}} 
				\put(24.5,24.0){\line(0,-1){22.5}}
				\put(24.5,24.0){\line(1,0){20.0}}
				\put(44.5,24.0){\line(0,-1){22.5}}
				\put(24.5,1.5){\line(1,0){20.0}}
				\put(50,35){\vector(-3,-1){15}}
				\put(51,35){$T$}
				\put(50,18){\vector(-3,-1){15}}
				\put(51,18){$T^\perp$}
			\end{picture}
		}
	\end{center}
	\caption{\label{fig:Delta} 
		The range of $\rho$ determines an orthogonal decomposition of the
		space of matrices as sketched in the figure. The space $T$ is the set of
		matrices $\sigma$ whose compression onto $\ker\rho$ vanishes
		(c.f.~Eq.~(\ref{eqn:T})). With respect to an eigenbasis of $\rho$,
		elements of $T$ are supported on the handle-shaped region shown
		above.
	}
\end{figure}

Following \cite{candes_exact_2009,candes_power_2009}, decompose
$\Delta=\Delta_T + \Delta_T^\bot$, with $\Delta_T \in T, \Delta_T^\bot
\in T^\bot$ (see Fig.~\ref{fig:Delta}). (The reason for doing this will become clear
momentarily).

The proof proceeds as follows
\begin{enumerate}
	\item
	In Section~\ref{sec:scales} we show that $\Delta$ is infeasible (fulfills
	(\ref{eqn:infeasible})) as soon as $\|\Delta_T\|_2$ is ``much
	larger'' than $\|\Delta_T^\bot\|$.

	\item
	The previous statement utilizes a large-deviation bound for
	operator-valued random variables, taken from \cite{ahlswede_strong_2002}. We
	repeat the proof of this powerful tool in
	Section~\ref{sec:ahlswede}.

	\item
	We go on to show that 
	\begin{equation*}
		\|\rho+\Delta\|_1 >
		\|\rho\|_1+\big(\sign\rho+\sign\Delta_T^\bot,\Delta\big)
	\end{equation*}
	in Section~\ref{sec:linearize}. Thus, as soon as the scalar product on the
	r.h.s.\ is positive, we conclude that $\Delta$ fulfills
	(\ref{eqn:tracenorm}). We then borrow a powerful idea from
	\cite{candes_exact_2009,candes_power_2009},
	employing a ``dual certificate''. More precisely it is shown that
	the aforementioned scalar product is guaranteed to be positive, as
	long as there is a matrix $Y\in\range\R$ such that ({\it i})
	$\P_T Y$ is close to $\sign\rho$, and ({\it ii}) $\|\P_T^\bot Y\|$ is
	small. 

	\item
	Section~\ref{sec:smallBases} establishes the existence of a certificate
	$Y$ in the case of bases with small operator norm. This is probably
	the most (comparatively) difficult part of the proof, and the one
	differing most from previous approaches.

	\item
	The construction of the previous section can be modified to work
	with any operator basis. Details are given in
	Section~Section~\ref{sec:generalBases}. This completes the proof of
	the main result.

	\item
	In Sections~\ref{sec:martingale},~\ref{sec:tight} we introduce some
	martingale techniques and put them to use to derive tighter bounds.

	\item
	Section~\ref{sec:nonHermitian} deals with non-Hermitian
	matrices.	
\end{enumerate}

Throughout, we will use the notation $m=n r \kappa$. The
``oversampling factor'' $\kappa$ describes the leverage we allow
ourselves by going beyond the minimum number of parameters needed to
describe $\rho$.

We use round parentheses
$(\sigma_1,\sigma_2)=\tr\sigma_1^\dagger\sigma_2$ for the
Hilbert-Schmidt inner product, and angle brackets
$\langle\psi,\phi\rangle$ for the standard inner product on $\CC^n$.

Let $s_i$ be the singular values of a matrix $\sigma$. The usual
matrix norms are
\begin{eqnarray*}
	\|\sigma\| &=& \max_i s_i, \\
	\|\sigma\|_2 &=& (\sigma,\sigma)^{1/2} = \left(\sum_i
	s_i^2\right)^{1/2},\\
	\|\sigma\|_1 &=& \tr|\sigma|=\sum_i s_i.
\end{eqnarray*}

Both the identity matrix and the identity function on more general
spaces are denoted by $\Id$.

We will frequently encounter inequalities between matrices, which are
understood in the usual sense: $\sigma_1 \leq \sigma_2$ if and only if
$\sigma_1 - \sigma_2$ is positive semi-definite (a convention
sometimes referred to as matrix order or L\"owner partial order).

As mentioned in the introduction (Section~\ref{sec:setting}), $\sign\sigma$ is the matrix
resulting from the application of the sign function to the
eigenvalues of $\sigma$.

\subsection{First case: large $\Delta_T$}\label{sec:scales}

In this section, we show that $\Delta$ is infeasible (with high
probability) if
$\Delta_T$ is much larger than $\Delta_T^\bot$.

If $\|\R\Delta_T\|_2 > \|\R\Delta_T^\bot\|_2$, then
\begin{equation*}
	\|\R\Delta\|_2 = \|\R\Delta_T +\R\Delta_T^\bot\|_2 
	\geq \|\R\Delta_T\|_2 - \|\R\Delta_T^\bot\|_2 > 0
\end{equation*}
To find criteria for this
situation to occur, we need to put a lower bound on $\|\R\Delta_T\|_2$
and an upper bound on $\|\R\Delta_T^\bot\|_2$. For the latter:
\begin{equation}\label{eqn:Rnorm}
	\|\R\Delta_T^\bot\|_2^2 
	= (\R\Delta_T^\bot, \R\Delta_T^\bot)
	\leq \|\R\|^2 \,\|\Delta_T^\bot\|_2^2.
\end{equation}
It's easy to see that $\|\R\|$ equals $n^2/m$ times the
highest number of collisions $C:=\max_i|\{j\,| A_i = A_j\}|$. This
number, in turn, is certainly smaller than $m$ (a truly risk-averse
estimate). All in all:
\begin{equation}\label{eqn:rdtb}
	\|\R\Delta_T^\bot\|_2\leq n^2\,\|\Delta_T^\bot\|_2.
\end{equation}
Likewise,
\begin{eqnarray}
	\|\R\Delta_T\|_2^2 
	&=& (\R\Delta_T, \R \Delta_T) \nonumber \\
	&\geq& \frac{n^2}{m} (\Delta_T, \R \Delta_T) 
	= \frac{n^2}{m} (\Delta_T,\P_T \R\P_T \Delta_T) \nonumber \\
	&\geq& \frac{n^2}{m} 
		\big(1- \|\P_T - \P_T \R \P_T\|\big)
	 \|\Delta_T\|_2^2.
	\label{eqn:A}
\end{eqnarray}
This makes $\P_T \R \P_T$ an object of interest.
Let $\P_{A_i}$ be the (matrix-valued) orthogonal projection onto
$w_{A_i}$. Then the identity
\begin{eqnarray*}
	\EE[\R]&=&\frac{n^2}{m} \sum_{i=1}^m
	\EE\big[\P_{A_i}\big]=\Id,\\
\end{eqnarray*}
follows directly from the fact that the matrices $\{w_a\}$ form an
ortho-normal basis by definition. We conclude that
$\EE[\P_T \R \P_T]=\P_T$. Thus, in order to evaluate
(\ref{eqn:A}), we need to bound the deviation of $\P_T \R \P_T$ from
its expectation value $\P_T$ in operator norm for small $m$.  In
\cite{candes_exact_2009}, this problem was treated using a bound known
as ``Rudelson selection principle'' \cite{rudelson_random_1999}. We will derive
a similar bound in the next section, as a corollary of the already
mentioned large-deviation theorem for matrix-valued random variables
from \cite{ahlswede_strong_2002}.  The result (proven in
Section~\ref{sec:ahlswede} below) reads:

\begin{lemma}\label{lem:Adev}
	It holds that
	\begin{equation}\label{eqn:Adev}
		\Pr[\|\P_T \R \P_T-\P_T\| \geq t ] \leq 4nr 
		\exp\left(-\frac{t^2 \kappa }{8 \nu }\right),
	\end{equation}
	for all $t<2$.
\end{lemma}

We assume in the following that (\ref{eqn:Adev}) holds with $t=1/2$.
Denote the probability of that event not occurring by $p_1$. 
(Many statements in this proof will hold only up to a small
probability of failure. We will defer an explicit calculation of these
failure probabilities until the very end of the argument, when all
parameters have been chosen).
Then, using (\ref{eqn:rdtb}), (\ref{eqn:A}), we have that $\R\Delta
\neq 0$ if
\begin{equation*}
	\frac{n^2}{2m} \|\Delta_T\|_2^2 \geq n^4 \|\Delta_T^\bot\|_2^2
	\Leftrightarrow
	\|\Delta_T\|_2^2 \geq 2mn^2\, \|\Delta_T^\bot\|_2^2.
\end{equation*}
For the next sections, it is thus sufficient to treat the case of 
\begin{equation}\label{eqn:scales}
	\|\Delta_T\|_2 < \sqrt{2m} n \,\|\Delta_T^\bot\|_2 <n^2
	\|\Delta_T^\bot\|_2.
\end{equation}

\emph{Remark:} Repeating the calculations in this section without the
trivial estimate $C<m$, the last coefficient in (\ref{eqn:scales}) can
be improved from $n^2$ to $\sqrt{\frac{2 C^2 n}{\kappa r}}$. Since $C$
is $O(\ln n)$ with very high probability, this would look like a
major improvement.  However, because only the logarithm of the
coefficient enters our final estimate of the number of samples
required, we will content ourselves with $n^2$ on the grounds that it
is a simpler expression.

\subsection{Operator large deviation bounds}
\label{sec:ahlswede}

The material in the first paragraph below is taken from
\cite{ahlswede_strong_2002}. We repeat the argument to make the
presentation self-contained. It is an elementary -- yet very powerful
-- large deviation bound for matrix-valued random variables.  The
basic recipe is this: take a textbook proof of Bernstein's inequality
and substitute all inequalities between real numbers by matrix
inequalities (in the sense of matrix order, see
Sec.~\ref{sec:notation}). 

We start by giving a basic Markov-inequality.  Let $\Theta$ be the
``operator step function'' defined by
\begin{equation*}
	\Theta(\sigma) = \left\{
		\begin{array}{ll}
			0 \quad & \sigma < \Id \\
			1 \quad & \sigma\not< \Id.
		\end{array}
	\right.
\end{equation*}
If $\sigma$ is
positive semi-definite, the trivial estimate $\Theta(\sigma)\leq \tr
\sigma$ holds. Thus, for any number $\lambda>0$ and matrix-valued
random variable $S$:
\begin{eqnarray}
	\Pr[S \not\leq t \Id ] 
	&=&\Pr[S -t \Id \not\leq 0] 
	=\Pr\big[e^{\lambda S- \lambda t\Id} \not\leq \Id \big] \nonumber \\
	&=&\EE\big[\Theta(e^{\lambda S-\lambda t\Id}) \big] 
	\leq \EE\big[\tr e^{\lambda S-\lambda t\Id}\big] \nonumber \\
	&=& e^{-\lambda t}\,\EE[\tr e^{\lambda S}]. \label{eqn:markov}
\end{eqnarray}
Now let $X$ be an operator-valued random variable, $X_i$ be
i.i.d.\ copies of $X$, and $S=\sum_{i}^m X_i$. Then
\begin{eqnarray}
	&&\EE\left[\tr \exp\bigg(\lambda\sum_i^m X_i \bigg)\right]\nonumber \\
	&\leq&
	\EE\left[\tr\exp\bigg(\lambda\sum_i^{m-1} X_i
	\bigg)\exp(\lambda X_m)\right]\nonumber \\
	&=&
	\tr\left(
	\EE[\exp\bigg(\lambda  \sum_i^{m-1} X_i
	\bigg)]\,\EE[ \exp(\lambda X)]\right)\nonumber \\
	&\leq&
	\EE\left[\tr \exp\bigg(\lambda \sum_i^{m-1} X_i \bigg)\right]
	\,
	\big\|\EE[ \exp(\lambda X)]\big\|\nonumber \\
	&\leq& \dots 
	\leq 
	\EE[\tr\exp(\lambda X_1)]
	\,
	\big\|\EE[ \exp(\lambda X)]\big\|^{m-1}\label{eqn:vorletzter}\\
	&\leq&	
	n\, \|\EE[ e^{\lambda X}]\|^m, 
	\label{eqn:momentPowers}
\end{eqnarray}
where the second line is the Golden-Thompson inequality
\cite{petz_survey_1994}.

Reference \cite{ahlswede_strong_2002} now goes on to derive a
Chernoff-Hoefding-type inequality for bounded $X_i\in[0,\Id]$. We
find it slightly more convenient to work with a Bernstein-type
estimate, bounding Eq.~(\ref{eqn:momentPowers}) by the second moments
of the $X_i$. (The derivation in the next paragraphs is influenced by
the proofs of the commutative version in \cite{dubhashi_concentration_2009,tao_additive_2006}).

Indeed, assume that $\EE[Y]=0$ and $\|Y\|\leq 1$ for some random
variable
$Y$. Recall the standard estimate
\begin{equation*}
	1+y \leq e^y \leq 1 + y + y^2 
\end{equation*}
valid for real numbers $y\in[-1,1]$ (and, strictly speaking, a bit
beyond).  From the upper bound, we get $e^Y \leq \Id + Y + Y^2$, as
both sides of the inequality are simultaneously diagonalizable. Taking
expectations and employing the lower bound:
\begin{eqnarray}
	\EE[e^Y] 
	\leq \Id + \EE[Y^2]
	\leq \exp(\EE[Y^2]), \label{eqn:Ysquared}
\end{eqnarray}
and thus $\|\EE[e^Y]\| \leq \|\exp(\EE[Y^2])\|=\exp(\|\EE[Y^2]\|)$.

These are all essential ingredients for the following theorem,
summarizing the results from this section.

\begin{theorem}[Operator-Bernstein inequality]\label{thm:bernstein}
	Let $X_i$, $i=1,\dots,m$ be i.i.d., zero-mean, Hermitian
	matrix-valued random
	variables. 
	Assume $V_0, c\in\RR$ are such that
	$\|\EE[X_i^2]\|\leq V_0^2$ and $\|X_i\|\leq c$.	
	Set $S=\sum_{i=1}^m X_i$ and let $V=m V_0^2$ (an upper bound
	to the variance of $S$).  Then
	\begin{equation}\label{eqn:bernstein}
		\Pr\big[\|S\|>t\big]
		\leq 2 n \exp\left(-\frac{t^2}{4V}\right),
	\end{equation}
	for $t \leq 2 V/c$, and
	\begin{equation}\label{eqn:bernstein2}
		\Pr\big[\|S\|>t\big]
		\leq 2 n \exp\left(-\frac{t}{2c}\right),
	\end{equation}
	for larger values of $t$. 
\end{theorem}

The second equation (\ref{eqn:bernstein2}) will be used only once, in
Section~\ref{sec:tight}. 

\begin{IEEEproof}
	Combine Eqs.~(\ref{eqn:markov}, \ref{eqn:momentPowers},
	\ref{eqn:Ysquared}) to get the estimate
	\begin{eqnarray*}
		\Pr[S\not\leq t \Id ]
		&\leq&
		n \exp\left(-\lambda t + \lambda^2mV_0^2\right).
	\end{eqnarray*}
	Let $s=t/V$ be the deviation in units of $V$. Then
	\begin{eqnarray*}
		\Pr[S\not\leq s V \Id ]
		&\leq&
		n \exp\left(-\lambda sV + \lambda^2V^2\right).
	\end{eqnarray*}
	Choose $\lambda=s/(2V)$. The exponent becomes
	\begin{equation*}
		- s^2/2 + s^2/4=-s^2/4
	\end{equation*}
	valid as long  as $\lambda\|X\|\leq 1$, which is certainly fulfilled
	if
	\begin{equation}\label{eqn:t}
		s \leq \frac{2V}{c}.
	\end{equation}
	If (\ref{eqn:t}) does not hold, set $\lambda=1/c$ and
	compute for the exponent
	\begin{eqnarray*}
		&&
		- sV/c + V^2/c^2
		=
		- sV/(2c) - 
		(
			sV/(2c)
			-V^2/c^2
		) \\
		&<&
		- sV/(2c)=-t/(2c).
	\end{eqnarray*}
	
	The same estimates hold  for $-S$, giving the advertised bound with
	the factor of $2$ coming from the union bound (which is also known as Boole's
	inequality: the probability of at least one of a set of events occurring
	is not larger than the sum of their individual probabilities).
\end{IEEEproof}

Note that for $n=1$, we recover the standard Bernstein inequality,
which we will also have the occasion to use.

We are in a position to supply the deferred proof of
Lemma~\ref{lem:Adev}. Recall that it was claimed that
\begin{equation*}
	\Pr[\|\P_T \R \P_T-\P_T\| \geq t ] \leq 4nr 
	\exp\left(-\frac{t^2 \kappa }{8 \nu }\right),
\end{equation*}
for all $t<2$.

\begin{IEEEproof}[Proof (of Lemma~\ref{lem:Adev})]
	For $a\in[1,n^2]$, let $\P_a$ be the orthogonal projection onto
	$w_a$.  We define a family of linear operators $Z_a$ by
	\begin{equation*}
		Z_a := \frac{n^2}{m}\,\P_T \P_a \P_T.
	\end{equation*}
	Then
	\begin{equation*}
		\P_T \R \P_T = \sum_{i=1}^m Z_{A_i}. 
	\end{equation*}
	Since $\EE[Z_{A_i}] = \frac1m \P_T$, the operator whose norm we want
	to bound can be written as
	\begin{equation*}
		\P_T \R \P_T - \P_T = \sum_{i=1}^m(Z_{A_i} - \EE[Z_{A_i}]). 
	\end{equation*}

	We will thus apply the Operator Bernstein inequality to the random
	variables
	$
		X_{A_i} := Z_{A_i} -\EE[Z_{A_i}].
	$
	To this end, we need to estimate the constants $V_0^2, c$  appearing
	in Theorem~\ref{thm:bernstein}. Compute:
	\begin{eqnarray*}
		\EE[Z_{A_i}^2]  
		=\frac{n^2}{m}
		\EE\big[(w_{A_i},\P_Tw_{A_i})\,Z_{A_i}\big].
	\end{eqnarray*}
	From Eq.~(\ref{eqn:coherence1}) we get $(w_{A_i},\P_Tw_{A_i})\leq
	\frac{2\nu r}{n}$ and thus
	\begin{eqnarray*}
		\EE[Z_{A_i}^2]  
		&\leq& \frac{n^2}{m} \frac{2\nu r}{n} 
		\EE\big[Z_{A_i}\big] = \frac{2n \nu r}{m^2}\,\P_T,
	\end{eqnarray*}
	having used that $Z_{A_i}\geq 0$ (matrix order). Hence
	\begin{eqnarray*}
		\big\|\EE[X_{A_i}^2] \big\|
		&=&
		\big\| \EE[Z_{A_i}^2]-\EE[Z_{A_i}]^2 \big\| \\
		&\leq& \frac{2 n \nu r -1}{m^2} \|\P_T\|
		\leq \frac{2 n \nu r}{m^2} = \frac{2 \nu }{m \kappa}=:V_0^2.
	\end{eqnarray*}
	
	Next:
	\begin{eqnarray*}
		\|X_{A_i}\| 
		&=& \frac1m \Big\|n^2\P_T \P_{A_i} \P_T - \P_T\Big\| \\
		&<& \frac1m\Big\|n^2\P_T \P_{A_i} \P_T\Big\| 
		= \frac{n^2}m \|\P_T w_{A_i}\|_2^2 \\
		&\leq& \frac{n^2}{m} 2 \nu \frac{r}{n} = \frac{2 \nu n r}{m} =
		\frac{2 \nu}{\kappa} =: c,
	\end{eqnarray*}
	so that
	\begin{equation*}
		2 m V_0^2 \frac1{\|X_{A_i}\|}
		\geq
		\frac{2 m \nu n r}{m^2} \frac{\kappa}{\nu}
		=
		\frac{2 \kappa n r}{m} = 2.
	\end{equation*}
	The claim follows from Theorem~\ref{thm:bernstein}.
\end{IEEEproof}

\subsection{Second case: small $\Delta_T$}\label{sec:second}
\label{sec:linearize}

In this section, we will show that 
\begin{eqnarray}
	\|\Delta_T\|_2 &<& n^2 \|\Delta_T^\bot\|_2,
	\label{eqn:scaleCondition} \\
	\Delta&\in&\range\R^\bot \label{eqn:rbot}
\end{eqnarray}
together imply $\|\rho+\Delta\|_1 > \|\rho\|_1$, \emph{if} we can find
a ``certificate'' $Y\in\range\R$ with certain properties.  The basic
line of argument is similar to the one given in Section 3 of
\cite{candes_exact_2009}. 

Set $U=\range\rho$ and let $P_U$ be the orthogonal projection
onto $U$.
We will make repeated use of the basic identity
\begin{equation*}
	\|\sigma\|_1= \tr|\sigma| = \tr((\sign \sigma)\sigma)=(\sign \sigma,
	\sigma)
\end{equation*}
(recall the definition of $\sign$ from Section~\ref{sec:setting}).
We then find
\begin{eqnarray}
	&&\|\rho+\Delta\|_1\nonumber \\
	&\geq&
	\|P_U(\rho+\Delta)P_U\|_1+
	\|P_U^\bot(\rho+\Delta)P_U^\bot\|_1 \label{eqn:pinching} \\
	&=&
	\|\rho+P_U \Delta P_U \|_1+
	\|\Delta_T^\bot\|_1 \nonumber\\
	&\geq&
	(\sign\rho,\rho+P_U \Delta P_U)+
	\big(\sign\Delta_T^\bot,\Delta_T^\bot\big) \label{eqn:duality} \\
	&=&
	\|\rho\|_1+
	(\sign\rho,P_U \Delta P_U)+
	(\sign\Delta_T^\bot,\Delta_T^\bot) \nonumber\\
	&=&
	\|\rho\|_1+\big(\sign\rho+\sign\Delta_T^\bot,\Delta\big).
	\label{eqn:subgradient}
\end{eqnarray}
The estimate (\ref{eqn:pinching}) is sometimes known as the ``pinching
inequality'' (\cite{bhatia_matrix_1997}, Problem II.5.4), and in line
(\ref{eqn:duality}) we used H\"older's inequality:
$
	(\sigma_1, \sigma_2) \leq \|\sigma_1\| \, \|\sigma_2\|_1.
$

To conclude that $\|\rho+\Delta\|_1 > \|\rho\|_1$, it is hence
sufficient to show that 
$
	(\sign \rho+\sign\Delta_T^\bot,\Delta)>0.
$
Choose any $Y\in\range\R$. Using (\ref{eqn:rbot}):
\begin{equation}\label{eqn:cond}
	(\sign \rho+\sign\Delta_T^\bot,\Delta)=
	\big(\sign\rho+\sign\Delta_T^\bot - Y,\Delta\big).
\end{equation}
Assume that $Y$ fulfills 
\begin{equation}\label{eqn:conds1}
	\|\P_T Y - \sign \rho \|_2 \leq \frac1{2n^2}, 
	\qquad
	\|\P_T^\bot Y \| \leq \frac12.
\end{equation}
Then (\ref{eqn:cond}) becomes
\begin{eqnarray}
	&&\big(\sign\rho+\sign\Delta_T^\bot - Y,\Delta\big)  \nonumber \\
	&=&
	\big(\sign\rho-Y, \Delta_T\big) + 
	\big(\sign\Delta_T^\bot - Y,\Delta_T^\bot\big) \nonumber \\
	&\geq&
	\frac12 \|\Delta_T^\bot\|_1 -
	\frac1{2n^2} \|\Delta_T\|_2 
	\geq
	\frac12 \|\Delta_T^\bot\|_2-
	\frac{1}{2n^2}\|\Delta_T\|_2  \nonumber \\
	&\geq&
	\frac14\|\Delta_T^\bot\|_2. \nonumber 
\end{eqnarray}

We summarize. Assume there is a certificate $Y\in\range\R$ fulfilling
(\ref{eqn:conds1}). Let $\sigma^\star$ be the solution of the
optimization problem, let $\Delta^\star=\rho-\sigma^\star$. Then
$\Delta^\star$ must fulfill (\ref{eqn:rbot}), for else it would be
unfeasible. It must also fulfill (\ref{eqn:scaleCondition}), by
Section~\ref{sec:scales}. But then, from the previous calculation
$(\Delta^\star)_T^\bot$ must be zero, as otherwise
$\|\sigma^\star\|_1>\|\rho\|_1$. This implies that $(\Delta^\star)_T$
is also zero, again using (\ref{eqn:scaleCondition}).  So
$\Delta^\star$ is zero, and therefore $\sigma^\star=\rho$ is the unique
solution to (\ref{eqn:optimizationR}).

It remains to prove the existence of the certificate $Y$.

\subsection{The certificate: bases of Fourier type}
\label{sec:smallBases}

In this section, we construct a $Y\in\range\R$ with
\begin{eqnarray}\label{eqn:conds}
	\|\P_T Y - \sign \rho \|_2 \leq \frac1{2n^2}, \qquad
	\|\P_T^\bot Y \| \leq \frac12
\end{eqnarray}
assuming that $\max_a\|w_a\|^2\leq \frac{\nu}{n}$. A modified proof
valid in the general case will be given in
Section~\ref{sec:generalBases}. In previous approaches to matrix
completion, this step was the most involved, covering dozens of pages.
We present a strongly simplified proof using two key ideas: a
further application of the operator Bernstein inequality; and a
certain, recursive random process which quickly converges to the
sought-for $Y$.

\subsubsection{Intuition}

A first, natural ansatz for finding $Y$ could be as follows. Define
\begin{equation}\label{eqn:failure}
	X_a = \frac{n^2}{m}\,w_a (w_a, \sign\rho),
	\qquad
	Y=\sum_i^m X_{A_i}.
\end{equation}
It is obvious that $Y$ is in the range of $\R$ and
that its expectation value (equal to $\sign\rho$) fulfills the
conditions in (\ref{eqn:conds}). What is more, the operator Chernoff
bound can be used to control the deviation of $Y$ from that expected
value -- so there is hope that we have found a solution. However, a
short calculation shows that convergence is (barely) too slow for our
purposes.

Intuitively, it is easy to see what is ``wrong'' with the previous
random process. Assume we sample $k<m$ basis elements. Employing
(\ref{eqn:failure}), our general ``best
guess'' at this point for a matrix $Y_1$ which resembles $\sign\rho$
on $T$ (i.e.\ with $\|\P_T Y_1-\sign\rho\|_2$ ``small'') would be
\begin{equation*}
	Y_1 = \frac{n^2}{k} \sum_i^k w_{A_i} (w_{A_i}, \sign\rho).
\end{equation*}
Now given this information, the matrix we really should be
approximating in the next steps is $\P_T(\sign\rho-Y_1)$. The process
(\ref{eqn:failure}), in contrast, does not update its ``future strategy
based on past results''.  Trying to perform better, we will draw a
further batch of $k$ coefficients and set
\begin{equation*}
	Y_2 = Y_1 + \frac{n^2}{k} \sum_{i=k+1}^{2k} w_{A_i} \big(w_{A_i},
	\sign\rho - \P_T Y_1\big).
\end{equation*}
The sequence $\P_T Y_i$ will be shown to converge exponentially fast
to $\sign\rho$. For reasons which should be all too obvious from
Fig.~\ref{fig:golfing}, we will call this adapted strategy the
\emph{golfing scheme}.

On the one hand, the size $k$ of the batches will have to be chosen
large enough to allow for the application of the operator
large-deviation bounds tailored for \emph{independent} random
variables. On the other hand, $k$ must not be too large, as the speed
of convergence is exponential in $l=m/k$. 

\subsubsection{Proof}

Before supplying the details of this scheme, we state a lemma which
will allow us to control the operator norm $\|\P_T^\bot Y\|$ of the
approximations. The operator-Bernstein inequality makes this, once
again, a simple calculation.

\begin{lemma}\label{lem:pbot1}
	Let $F\in T$. 
	Then
	\begin{eqnarray*}
		\Pr\bigg[ \big\|\P_T^\bot \R F\big\| > t 
		\bigg]
		&\leq& 
		2n \exp\left(-\frac{t^2 \kappa r}{4 \nu \|F\|_2^2}\right) 
	\end{eqnarray*}
	for $t\leq \sqrt{2/r} \|F\|_2$, and
	\begin{eqnarray*}
		\Pr\bigg[ \big\|\P_T^\bot \R F\big\| > t 
		\bigg]
		&\leq& 
		2n \exp\left(-\frac{t \sqrt r\kappa }{2\sqrt 2 \nu \|F\|_2}\right)
	\end{eqnarray*}
	for larger values of $t$.
\end{lemma}

\begin{IEEEproof}
	It suffices to treat the case where $\|F\|_2=1$. Set
	\begin{equation*}
		X_a = \frac{n^2}{m}  \P_T^\bot w_a\,	(w_a, F).
	\end{equation*}
	Then $\sum_{i}^m X_{A_i}=\P_T^\bot \R F$, and
	\begin{equation*}
		\EE[X_{A_i}] = \frac{1}{m} \P_T^\bot F = 0.
	\end{equation*}
	Using (\ref{eqn:coherenceNorm}) and the fact that $\|\P_T^\bot
	w_a\|\leq \|w_a\|$ we estimate the variance:
	\begin{eqnarray}
		\|\EE[X_{A_i}^2]\| 
		&\leq& \frac{n^2}{m^2} \sum_a (w_a, F)^2 \|(\P_T^\bot w_a)^2\| 
		\label{eqn:operatorNormEssential} \\
		&\leq& \frac{n^2}{m^2} \frac{\nu}{2n} \|F\|_2^2 
		= \frac{n \nu}{m^2}
		= \frac{\nu}{m \kappa r}
		:=V_0^2. \nonumber
	\end{eqnarray}
	Next,
	\begin{eqnarray*}
		\|X_{A_i}\|\leq \frac{n^2}{m} \sqrt{\frac{\nu}{n} \frac{2\nu r}{n}}
		=
		\frac{n\nu \sqrt {2r}}{m}
		=\frac{\sqrt 2 \nu}{\sqrt {r} \kappa},
	\end{eqnarray*}
	so that
	\begin{equation*}
		2mV_0^2/\|X_{A_i}\| 
		\geq
		\frac{2m \nu}{m\kappa r} \frac{\sqrt{r}\kappa}{\sqrt 2\nu }
		=
		\frac{\sqrt 2}{\sqrt r}.
	\end{equation*}
	Now use Theorem~\ref{thm:bernstein}.
\end{IEEEproof}

We sample $l$ batches of basis elements, the $i$th set
consisting of $m_i=\kappa_i r n$ matrices. 
\begin{figure}
	\begin{center}
		\includegraphics[scale=1.3]{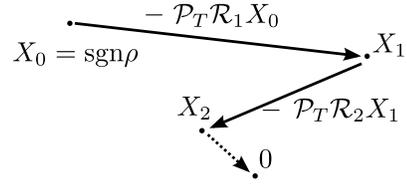}
	\end{center}
	\caption{\label{fig:golfing} Caricature of the ``golfing scheme''
	used to construct the certificate. In the $i$th step,
	$X_{i-1}$ designates the vector we aim to represent. The
	approximation of $X_{i-1}$ actually obtained is $\P_T \R_i X_{i-1}$.
	The distance of the new goal $X_i = X_{i-1} - \P_T \R_i X_{i-1}$ 
	to the origin is guaranteed to be only half the previous one.	
	The
	sequence $X_i$ thus converges exponentially fast to the origin.  }
\end{figure}

For $1\leq i \leq l$, let
\begin{equation*}
	\R_i: \sigma \mapsto \frac{n^2}{m_i}
	\sum_{j=m_1+\dots+m_{i-1}+1}^{m_1+\dots+m_i} w_{A_j}\,
	(w_{A_j},\sigma)
\end{equation*}
be the sampling operator associated with the $i$th batch and set
\begin{eqnarray*}
  X_0 = \sign \rho, \qquad
	Y_i = \sum_{j=1}^i \R_j X_{j-1}, \qquad
	X_i = \sign\rho-\P_T Y_i
\end{eqnarray*}
(see Fig.~\ref{fig:golfing}). From this, we get
\begin{eqnarray}\label{eqn:notneumann}
	&&X_i = \\ 
	&&(\Id - \P_T \R_i \P_T) 
	(\Id - \P_T \R_{i-1} \P_T) 
	\dots
	(\Id - \P_T \R_1 \P_T) X_0.
	\nonumber
\end{eqnarray}

Assume that in the $i$th run
\begin{equation}\label{eqn:assumption1}
	\|(\Id - \P_T \R_i\P_T )X_{i-1}\|_2<c_i \|X_{i-1}\|_2.
\end{equation}
Denote the probability of this event not occurring by
$p_2(i)$ (recall that $p_1$ has been defined in
Section~\ref{sec:scales}). Clearly, if (\ref{eqn:assumption1}) does
hold for all $i$, then
\begin{eqnarray*}
	\|X_i\|_2
	= \|(\Id - \P_T\R_i\P_T)  X_{i-1}\|_2 
	\leq c_i \|X_{i-1}\|_2,
\end{eqnarray*}
so that $\| X_i\|_2 \leq \sqrt r \prod_{j=1}^i c_j$.

Assume further that for all $i$ the estimate
\begin{equation*}
	\|\P_T^\bot \R_i X_{i-1}\| \leq t_i \|X_{i-1}\|_2
\end{equation*}
is true, with $p_3(i)$ bounding the probability of failure. 

Then
\begin{eqnarray*}
	\|\P_T^\bot Y_l\| 
	\leq
	\sum_{i=1}^l \|\P_T^\bot \R_j X_{j-1}\|
	\leq
	\sum_{i=1}^l t_i \|X_{i-1}\|_2.
\end{eqnarray*}

A first simple choice of parameters (to be refined in
Section~\ref{sec:tight}) is
\begin{eqnarray*}
	c_i &=& 1/2, \\
	t_i &=& 1/(4\sqrt r), \\
	\kappa_i &=& 64 \nu(\ln(4nr)+\ln(2l) + \beta \ln n)
\end{eqnarray*}
for some $\beta>0$. It follows that
\begin{eqnarray*}
	\|X_i\|_2 \leq \sqrt r 2^{-i}, \qquad
	\|\P_T^\bot Y_l\|\leq
	\frac14 \sum_{i=1}^l 2^{-(i-1)} < \frac12.
\end{eqnarray*}
With $l=\lceil \log_2(2 n^2 \sqrt r) \rceil $, the conditions in
Equation~(\ref{eqn:conds}) are met. Using Lemma~\ref{lem:Adev} and
Lemma~\ref{lem:pbot1} the failure probabilities become
\begin{eqnarray*}
	p_1 &\leq& 4nr\exp\left(-\frac{\kappa}{32\nu}\right), \\
	p_2(i)&\leq& 4nr\exp\left(-\frac{\kappa_i}{32\nu}\right), \\
	p_3(i)&\leq& 
	2n \exp\left(-\frac{\kappa_i}{64  \nu }\right)
\end{eqnarray*}
all of which are bounded above by $\frac 1{2l} n^{-\beta}$.
Theorem~\ref{thm:main}
for Fourier-type bases thus follows from a simple application of the
union bound. The number of coefficients sampled must exceed
\begin{eqnarray*}
	&&m\\
	&=& 
	l\kappa_i = 64 \nu(\ln(4nr)+\ln(2l)+\beta\ln
	n)\log_2(2n^2\sqrt r)rn \\ 
	&=& O(r n \nu(1+\beta) \ln^2n).
\end{eqnarray*}

\subsubsection{Discussion}
\label{sec:discussion}

The ``golfing scheme'' above could be described as a ``sequential''
way of building the certificate vector: every time we sample a basis
element $w_a$, we assign a coefficient $c_a=(w_a, X_i)$ to it, but
never alter our previous choices. This contrasts with the more
``holistic'' method employed in
\cite{candes_exact_2009,candes_power_2009}, where $Y$ was constructed
by directly inverting $\P_T \R \P_T$: 
\begin{equation}\label{eqn:inverting}
	Y=\R P_T (P_T \R P_T)^{-1} \sign\rho.
\end{equation}

Presumably, the most \emph{optimal} sequential scheme is the one which
chooses the coefficient $c_a$ in every step such as to minimize the
distance to the vector we aim to approach. If the distance is measured
in 2-norm, it is simple to write down a closed-form expression for
that choice. However, such a strategy introduces strong dependencies
into the random process, which make an analysis challenging.  The
elementary i.i.d.\ tools
employed in this paper are no longer applicable. This intuition
motivates considering \emph{martingale generalizations} of the
operator-large deviation bounds of \cite{ahlswede_strong_2002}. We will indeed
prove a deviation estimate for matrix-valued martingales in
Section~\ref{sec:martingale}. Whether this bound is sufficient to
analyze the ``optimal sequential scheme'' remains unclear.

Another observation is that, since Lemma~\ref{lem:Adev} provides a
\emph{uniform} bound on $\|(\P_T\R\P_T - \Id)X\|_2$, there is no need
for the iterative scheme to chose a \emph{different} set of basis
elements in each run, in order to achieve exponential convergence of
$\P_T Y_i \to \sign \rho$. Iterating over a single fixed set of $O(n r \ln
n)$ basis elements would equally do the job. Unfortunately, the
statement of Lemma~\ref{lem:pbot1} is not uniform in $F\in T$,
necessitating the less-optimal approach used above in order to control
$\|\P_T^\bot Y_i\|$.  However, a smart substitute for the crude union
bound could potentially remedy this situation.

By the same token, one can replace Lemma~\ref{lem:Adev} by a
non-uniform estimate. The golfing scheme only requires that
$\|(\P_T\R_i\P_T - \Id)X_i\|_2$ be small, which is much easier to
guarantee than a similar bound on $\|\P_T\R_i\P_T-\P_T\|$.  This is
precisely the role of Theorem~\ref{thm:vecBernstein} below, on which
bounds of order $O(r n \nu \ln n)$ can be based (see
Section~\ref{sec:tight}).

We remark that \cite{candes_exact_2009,candes_power_2009} analyzed
(\ref{eqn:inverting}) by expanding the inverse into a Neumann series
\begin{equation}\label{eqn:neumann}
	(\P_T \R\P_T)^{-1} = \sum_{n=0}^\infty (\Id - \P_T\R\P_T)^n.
\end{equation}
There is a formal analogy between this series and our
construction, in particular in the light of (\ref{eqn:notneumann}).
Note however, that the product in (\ref{eqn:notneumann}) involves
distinct and independently drawn sampling operators $\R_i$ in every
factor. Informally speaking, this added degree of independence seems
to make (\ref{eqn:notneumann}) a more benign object than the powers
$(\Id-\P_T \R\P_T)^n$ in (\ref{eqn:neumann}).

\subsection{The certificate: general case}
\label{sec:generalBases}

In this section, we show that the construction of $Y$ described above
continues to work if the assumption (\ref{eqn:coherenceNorm}) on the
operator norm of the basis elements is replaced by the incoherence
properties (\ref{eqn:coherence1}, \ref{eqn:coherence2}).

Indeed, in the discussion of the golfing scheme, we referred to the
operator norm of $w_a$ exactly once.  In the proof of
Lemma~\ref{lem:pbot1}, we considered the quantity
\begin{equation}\label{eqn:generalRV}
		X_a = \frac{n^2}{m}  \P_T^\bot w_a\,	(w_a, F).
\end{equation}
After 
Equation~(\ref{eqn:operatorNormEssential}), the variance
\begin{equation*}
		\|\EE[X_{A_i}^2]\| 
		\leq \frac{n^2}{m^2} \sum_a (w_a, F)^2 \|(\P_T^\bot w_a)^2\| 
\end{equation*}
was upper-bounded using the fact that $\|(\P_T^\bot w_a)^2\|\leq
\frac{\nu}{n}$. Clearly the absence of this assumption can be
compensated for by a suitable bound on $(w_a, F)^2$. This will be made
precise below.

Assume that $F$ is some matrix in $T$ with $\|F\|_2=1$. Further,
assume that at least one of the following two bounds
\begin{eqnarray}
	\max_a \|w_a\|^2 &\leq& \frac {\nu}{n}, \\
	\max_a \,|(w_a, F)|^2 &\leq& \frac{\nu}{n^2}.
\end{eqnarray}
holds.

Note that
\begin{eqnarray}\label{eqn:pq}
	\|\EE[X_{A_i}^2]\|
	&\leq&
	\frac{n^3}{m^2} 
	\max_\psi	
	\sum_a
		\big(w_a,F\big)^2 \,
		\frac 1n
		\langle \psi, w_a^2 \psi\rangle,
\end{eqnarray}
where the maximum is over all normalized vectors $\psi\in
(\range\rho)^\bot$. Let $\psi_0$ be a vector achieving the maximum.
Define two vectors $p,q$ in $\RR^{n^2}$ by setting their components to
\begin{equation}\label{eqn:dualityConds}
	q_a := \big(w_a,F\big)^2, \qquad 
	p_a:= \frac 1n \langle \psi_0, w_a^2 \psi_0\rangle
\end{equation}
respectively.
The assumption that $\|F\|_2^2=1$ implies that $\|q\|_1=\sum_a |q_a|=1$.
Slightly less obvious is the fact that the same is true for the other
vector: $\|p\|_1=1$, regardless of the basis chosen. This relation is
ascertained by the next lemma.

\begin{lemma}\label{lem:basisSquared}
	Let $\{w_a\}$, be a set of $n\times n$-matrices (not necessarily
	Hermitian) that fulfill the
	completeness
	relation
	\begin{equation}\label{eqn:completeness}
		\sum_a (\bar w_a)_{i_1, j_1} (w_a)_{i_2, j_2}
		= \delta_{i_1, i_2}\,\delta_{i_2, j_2}.
	\end{equation}
	Then
	\begin{equation*}
		\sum_a w_a^\dagger w_a = n\,\Id.
	\end{equation*}
\end{lemma}

\begin{IEEEproof}
	Compute:
	\begin{eqnarray*}
		\big(\sum_a w_a^\dagger w_a \big)_{i,j}
		= \sum_{a,k} (\bar w_a)_{k,i}  (w_a)_{k,j}
		= \sum_{k} \delta_{i, j} = n\,\delta_{i,j}.
	\end{eqnarray*}
\end{IEEEproof}

Thus,
\begin{equation*}
	\|p\|_1 = \sum_a p_a= \frac1n \langle\psi_0, n\Id \psi_0\rangle = 1.
\end{equation*}

We return to the vectors in (\ref{eqn:dualityConds}). The assumptions made
imply that at least one of the vectors is element-wise bounded above
by $\frac{\nu}{n^2}$. Thus
\begin{equation}\label{eqn:symmetry}
	\left|\sum_a p_a q_a \right|
	\leq \min\{\|p\|_1 \|q\|_\infty, \|p\|_\infty \|q\|_1\}	
	\leq \frac{\nu}{n^2}.
\end{equation}
Plugging this estimate into the computation of the variance
(\ref{eqn:pq}) we
obtain
\begin{eqnarray*}
	\|\EE[X_{A_i}^2]\|
	&\leq&
	\frac{n^3}{m^2} \frac{\nu }{n^2} = \frac{\nu}{m \kappa r}.
\end{eqnarray*}

We have proved the general analogue of Lemma~\ref{lem:pbot1}:

\begin{lemma}\label{lem:pbot2}
	Let $F\in T$. 
	Let $f\geq \|F\|_2$ be an upper bound on the 2-norm of $F$. 
	Assume that one of the two bounds
	\begin{eqnarray}
		\max_a \|w_a\|^2 &\leq& \frac {\nu}{ n}, \\
		\max_a \,(w_a, F)^2 &\leq& \frac{\nu}{n^2} f^2 
	\end{eqnarray}
	holds. Then
	\begin{eqnarray}
		\Pr\bigg[ \big\|\P_T^\bot \R F\big\| > t 
		\bigg]
		&\leq& 
		2n \exp\left(-\frac{t^2 \kappa r}{4 \nu f^2}\right), 
	\end{eqnarray}
	for $t\leq \sqrt{2/r} f$.
\end{lemma}

Next, we have to justify the bounds on $(w_a,F)^2$ we imposed in the
previous lemma. By assumption (\ref{eqn:coherence2}), the estimate
does hold  
for $F=\sign\rho$, i.e.\ Lemma~\ref{lem:pbot2} may be
applied during the first leg $X_0=\sign\rho$ of the ``golfing
scheme''. However, there is no a priori reason that the same be true
for $X_1=(\Id-\P_T\R_1\P_T) X_0$. For now, all we know about $X_1$
is that it is an element of $T$ and hence low-rank. This property
was enough for Fourier-type bases, but in the general case, it proves
too weak. We thus have to ensure that ``inhomogeneity'' of $X_i$
implies inhomogeneity of $X_{i+1}$, a fact that can be ascertained
using yet another Chernoff bound.

Let $\mu(F)=\max_a (w_a, F)^2$ be the maximal squared overlap between
$F$ and any element of the operator basis.

\begin{lemma}\label{lem:stayIncoherent}
	Let $F\in T$. Then
	\begin{equation*}
		\Pr\left[\mu\big((\Id-\P_T\R\P_T)F\big) > t \right]
		\leq
		2n^2 \exp\left(-\frac{t \kappa }{4 \mu(F) \nu}\right), 
	\end{equation*}
	for all  $t\leq \mu(F)$.
\end{lemma}

\begin{IEEEproof}
	Fix $b\in[1,n^2]$. Define 
	\begin{equation}\label{eqn:overlapsrv}
		X_a = \frac1m (w_b, F) - (w_b, \frac{n^2}{m} \P_T w_{a}) (w_{a}, F).
	\end{equation}
	Then
	\begin{equation*}
		\sum_i^m X_{A_i} 
		= \big(w_b, (\Id-\P_T\R\P_T) F).
	\end{equation*}
	Note that the first term in (\ref{eqn:overlapsrv}) is the
	expectation value of the second one. Therefore, $\EE[X_{A_i}]=0$ and
	the variance of $X_{A_i}$ is bounded above by the variance of
	the second term alone (as in the proof of Lemma~\ref{lem:Adev}):
	\begin{eqnarray*}
		\EE[X_{A_i}^2] 
		&\leq&
		\frac1{n^2} \sum_a (w_b, \frac{n^2}{m} \P_T w_a)^2 (w_a, F)^2 \\
		&\leq&
		\frac{n^2}{m^2} \mu(F) \sum_a (\P_T w_b, w_a)^2 \\
		&=&
		\frac{n^2}{m^2} \mu(F) \|\P_T w_b\|_2^2  \\
		&\leq& 
		\frac{n^2 \mu(F) \nu r}{m^2 n}
		=
		\frac{\mu(F) \nu}{m\kappa}=:V_0^2.
	\end{eqnarray*}

	Further,
	\begin{equation*}
		|X_{A_i}|\leq 
		\frac1m \mu(F)^{1/2}
		\left(
			1 + 
			n^2\frac{\nu r}{n}
		\right)
		=
		\frac1m \mu(F)^{1/2}(1+n\nu r).
	\end{equation*}

	Thus, from the Chernoff bound:
	\begin{eqnarray*}
		&&\Pr\bigg[|(w_b,(\Id - \P_T\R\P_T)F)| > \sqrt{t}\,
		\bigg]  \\
		\leq 
		&&2 \exp\left(-\frac{t\kappa}{4 \mu(F) \nu}\right)
	\end{eqnarray*}
	as long as $\sqrt t$ does not exceed
	\begin{eqnarray*}
		2 m V_0^2/|X_{A_i}|
		=
		\frac{2m \mu(F) \nu}{m\kappa} \frac{m}{\mu(F)^{1/2}(1+n\nu r)}
		\geq
		\mu(F)^{1/2}. 
	\end{eqnarray*}

	The advertised estimate follows by taking squares and applying the
	union bound over the $n^2$ elements of the basis.
\end{IEEEproof}

With these preparations made, we can repeat the ``golfing'' argument
from the last section. As an additional constraint, we demand that 
\begin{equation*}
	\mu(X_i) \leq c_i^2\,\mu(X_{i-1})
\end{equation*}
be fulfilled for all $i$, with probability of failure given by
$p_4(i)$.

Then, with
\begin{eqnarray*}
	c_i &=& 1/2, \\
	t_i &=& 1/(2\sqrt r), \\
	\kappa_i &=& 64 \nu(\ln(4n^2)+\ln(3l) + \beta \ln n) 
\end{eqnarray*}
it follows that
\begin{eqnarray*}
	\|X_i\|_2 &\leq& 2^{-i} \|\sign\rho\|_2 = 2^{-i} \sqrt r,  \\
	\mu(X_i) &\leq& 2^{-2i} \mu(\sign\rho) 
	\leq 
	 \frac{\nu}{n^2}\,(2^{-2i} r).
\end{eqnarray*}
Thus, in $i$th iteration of the golfing scheme, we can apply
Lemma~\ref{lem:pbot2} with $F=X_i$ and $f=2^{-i} \sqrt r$.

The failure probabilities $p_1, p_2(i)$ and $p_3(i)$ are as before.
Further
\begin{eqnarray*}
	p_4(i)&\leq& 2n^2\exp\left(-\frac{\kappa_i}{16\nu}\right),
\end{eqnarray*}
which, as the other probabilities, is bounded above by
$\frac 1{3l} n^{-\beta}$. By the union bound, Theorem~\ref{thm:main}
holds as long as 
\begin{equation*}
	m> \log_2(2n^2\sqrt r) 64 \nu(\ln(4n^2)+\ln(3l) + \beta \ln n)rn.
\end{equation*}

\section{Refined methods and generalizations}

\subsection{Martingale methods for matrix-valued random variables}
\label{sec:martingale}

The purpose of this section is two-fold. First, we derive a
dimension-free bound for the norm of the sum of vector-valued random
variables (Theorem~\ref{thm:vecBernstein}). Substituting
Lemma~\ref{lem:Adev} by this dimension-free analogue will enable us to
give tighter bounds of matrix recovery in Section~\ref{sec:tight} (see
discussion in Section~\ref{sec:discussion}). Such dimension-free bounds
for sums of vectors are
well-known \cite{ledoux_probability_1991} and we could in principle content
ourselves with citing an existing version. Making the proof explicit,
however, ensures that this document remains self-contained and allows
us to record a corollary which may be of independent interest. Indeed,
the simplest argument in \cite{ledoux_probability_1991} relies on a standard
large-deviation bound for real-valued martingales. We use the
occasion to prove an operator version (Theorem~\ref{thm:martingale})
of this martingale estimate, which generalizes the operator Chernoff
bound. This constitutes the second purpose of the present section.

Let $X_1, \dots, X_m$ be a sequence of random variables. We will use
the bold-face symbol $\Xb_{i}$ to refer to the set $\{X_1, \dots,
X_i\}$ of the first $i$ of these variables.
Theorem~\ref{thm:martingale} is an almost verbatim translation of the
real-valued statement in \cite{dubhashi_concentration_2009} (see also
\cite{mcdiarmid_concentration_1998}). To lift it to operator-valued
variables, we use exactly the same tricks that were employed in
\cite{ahlswede_strong_2002} to obtain the operator Chernoff bound
(c.f.\ our exposition in Section~\ref{sec:ahlswede}).

\begin{theorem}[Variance bound for matrix-valued martingales]
\label{thm:martingale}
	Let $X_0, \dots, X_m$ be arbitrary random variables.
	Let $Z_0=0$ and let $Z_1, \dots, Z_m$ be a sequence of ($n\times
	n$)-Hermitian matrix-valued random variables. Assume the martingale
	condition
	\begin{equation*}
		\EE[Z_i \,|\, \Xb_{i-1}] = Z_{i-1}
	\end{equation*}
	holds for $i=1,\dots,m$. Assume further that the martingale
	difference sequence
	$D_i = Z_i - Z_{i-1}$ respects
	\begin{equation*}
		\|D_i\|\leq c_i, \qquad
		\|\EE[D_i^2\,|\,\Xb_{i-1}]\|\leq \sigma_i^2.
	\end{equation*}
	Then, with $V=\sum_i^m \sigma_i^2$,
	\begin{equation}\label{eqn:martingale}
		\Pr\big[\|Z_m\|>t\big]
		\leq 2 n \exp\left(-\frac{t^2}{4 V}\right),
	\end{equation}
	for any $t \leq 2 V/(\max_i c_i)$.
\end{theorem}
	
\begin{IEEEproof}
	As in Section~\ref{sec:ahlswede}, 
	\begin{equation*}
		\Pr[S \not\leq t] \leq e^{-\lambda} \EE[\tr e^{\lambda Z_n}]
	\end{equation*}
	for any $\lambda>0$. Using Golden-Thompson:
	\begin{eqnarray}
		\EE[\tr e^{\lambda Z_m} ]
		&=&
		\EE\big[\EE[\tr e^{\lambda(Z_{m-1}+D_m)}|\Xb_{m-1}]\big] \nonumber\\
		&\leq&
		\EE\big[\EE[\tr e^{\lambda Z_{m-1}} e^{\lambda D_m}|\Xb_{m-1}]\big] 
		\nonumber\\
		&=&
		\EE\big[\tr e^{\lambda Z_{m-1}} \EE[e^{\lambda D_m}|\Xb_{m-1}]\big]
		\nonumber\\
		&\leq&
		\EE\big[\tr e^{\lambda Z_{m-1}} \|\EE[e^{\lambda
		D_m}|\Xb_{m-1}]\|\big].
		\nonumber
	\end{eqnarray}
	From the martingale condition:
	\begin{equation*}
		\EE[\lambda D_i|\Xb_{i-1}]
		=\lambda \EE[Z_{i}-Z_{i-1}|\Xb_{i-1}] 
		=0.
	\end{equation*}
	Once more, we will make use of the estimate $1+y \leq e^y \leq 1 + y
	+ y^2$ valid for $|y|\leq 1$: 
	\begin{eqnarray*}
		\EE[e^{\lambda D_i}| \Xb_{i-1} ]
		&\leq&
		\Id + \EE[\lambda D_i | \Xb_{i-1}]+ \EE[\lambda^2 D_i^2|\Xb_{i-1}] \\
		&=&
		\Id + \lambda^2 \EE[D_i^2|\Xb_{i-1}] \\
		&\leq&
		\exp\left(\lambda^2 \EE[D_i^2|\Xb_{i-1}] \right),
	\end{eqnarray*}
	as long as $\lambda \|D_i\|\leq 1$. Thus
	\begin{equation*}	
		\left\|
			\EE[e^{\lambda D_i}| \Xb_{i-1} ]
		\right\|
		\leq 
		\left\|
			\exp\left(\lambda^2 \EE[D_i^2|\Xb_{i-1}] \right)
		\right\|
		=	
		e^{\lambda^2 \sigma_i^2}.
	\end{equation*}
	By induction
	\begin{equation*}
		\EE[\tr e^{\lambda Z_n} ] 
		\leq
		\EE[\tr e^{\lambda Z_1}]
		e^{\lambda^2(\sigma_2^2+\dots+\sigma_n^2)}
		\leq
		n e^{\lambda^2 V}.
	\end{equation*}
	
	The claim follows by setting $\lambda=t/2V$.
\end{IEEEproof}

The next theorem is essentially contained in Chapter~6 of
\cite{ledoux_probability_1991} (see also
\cite{yurinskii_exponential_1974}). To keep the presentation
self-contained, we give a short proof in
Appendix~\ref{sec:vectorMartingale}.

\begin{theorem}[Vector Bernstein inequality]\label{thm:vecBernstein}
	Let $X_1,\dots,X_m$ be independent zero-mean vector-valued
	random variables.  Let
	\begin{equation*}
		N = \Big\|\sum_{i=1}^m X_{i}\Big\|_2.
	\end{equation*}
	Then
	\begin{equation*}
		\Pr\left[
			N \geq \sqrt V+ t
		\right]
		\leq  \exp\left(-\frac{t^2}{4 V}\right),
	\end{equation*}
	where $V=\sum_i \EE[\|X_i\|_2^2]$ and $t\leq V/(\max \|X_i\|_2)$.
\end{theorem}

We can now prove a non-uniform, but dimension independent version of
Lemma~\ref{lem:Adev}.

\begin{lemma}\label{lem:dimensionFree}
	Let $F\in T$. Then
	\begin{equation*}
		\Pr
		\left[
			\|(\P_T \R-\Id)F\|_2 \geq t\|F\|_2	
		\right] 
		\leq 
		\exp\left(-\frac{(t-\sqrt{2\nu/\kappa})^2 \kappa }{8 \nu }\right),
	\end{equation*}
	provided $t\leq 2/3$.
\end{lemma}

\begin{IEEEproof}
	Let 
	\begin{equation*}
		X_i = \frac{n^2}{m} \P_T w_{A_i} (w_{A_i},F) - \frac1m F.
	\end{equation*}
	Then
	\begin{eqnarray*}
		\EE[\|X_i\|_2^2] 
		&\leq&
		\frac{n^4}{m^2} \EE[(w_A,F)^2 \|\P_T w_A\|_2^2] \\
		&\leq&
		\frac{n^2}{m^2} 2\frac{\nu r}{n} \|F\|_2^2
		=\frac{2\nu}{m \kappa} \|F\|_2^2=:V_0^2.
	\end{eqnarray*}
	Next,
	\begin{eqnarray*}
		\|X_i\|_2
		&\leq& 
		\frac{\|F\|_2}{m} 
		\left(
			n^2\frac{2\nu r}{n} + 1
		\right)
		\leq  \frac{3 \nu \|F\|_2}{\kappa}.
	\end{eqnarray*}
	So that
	\begin{equation*}
		2V/\|X_i\|_2 \geq 2/3 \|F\|_2.
	\end{equation*}
	Now use Theorem~\ref{thm:vecBernstein}.
\end{IEEEproof}

\emph{Note added}: After the pre-print version of this paper was
published, the author was made aware of a related matrix-valued
martingale bound in \cite{christofides_expansion_2007}. The
derivations used in \cite{christofides_expansion_2007} are very
similar in spirit to ours (however, their results cannot be applied
directly to the problem treated here, because no variance information
is incorporated). A few months after our pre-print appeared, more
sophisticated matrix-valued martingale bounds were established in
\cite{tropp_user-friendly_2010}.

\subsection{Tighter bounds for Fourier-type bases}
\label{sec:tight}

We present a refined analysis of the ``golfing scheme'', which
achieves fairly tight bounds for Fourier-type bases. Compared to
Section~\ref{sec:smallBases}, there are two changes in the argument.
First, we use the dimension-free large deviation bound for vectors
derived in the previous section. Second, the parameters of the random
process used to construct the certificate are chosen more carefully.

Let $\alpha>4$ be a number to be chosen later.
We will analyze the following set of parameters for the golfing scheme:
\begin{eqnarray*}
	\kappa_i &=& 18(\ln \alpha+\beta)\nu c_i^{-2},  \\
	c_1&=&c_2=\frac{1}{2\ln^{1/2} n},\\
	c_i &=& \frac12\qquad (2<i\leq l),\\
	t_1&=&t_2=\frac1{4\sqrt r},\\
	t_i &=& \frac{\ln n}{4\sqrt r}\qquad (2<i\leq l), \\
	l&=&\lceil \log_2(2 n^2 \sqrt r) \rceil.
\end{eqnarray*}

Using the arguments from Section~\ref{sec:smallBases},
\begin{equation*}
	\| X_i\|_2 \leq \sqrt r \prod_{j=1}^i c_j
	=
	\sqrt r\, 2^{-i}
	\left\{
		\begin{array}{ll}
				\ln^{-i/2}n \qquad & i=1,2 \\
				\ln^{-1} n  \qquad & i>2. 
		\end{array}
	\right.
\end{equation*}
Thus
\begin{eqnarray*}
	\|\P_T^\bot Y_l\| 
	&\leq&
	\sum_{i=1}^l t_i \|X_{i-1}\|_2\\
	&\leq&
	\frac 14 \left(
		1 + 
		\frac12 \frac1{\ln^{1/2} n} +
		\frac14 \frac{\ln n}{\ln n} +
		\frac18 \frac{\ln n}{\ln n} + \dots
	\right)\\
	&<& \frac12,
\end{eqnarray*}
and
\begin{eqnarray*}
	\|X_l\|_2&=&\|\P_T Y_l - \sign\rho\|_2 \leq \frac1{2n^2}
\end{eqnarray*}
as required by (\ref{eqn:conds}). 

We look at the failure probabilities.
To bound $p_2(i)$, we make use of the
dimension-free estimate provided by Lemma~\ref{lem:dimensionFree}:
\begin{eqnarray*}
	p_2(i)
	&\leq&
	\exp\left(-
		\frac{
			(\frac 23 c_i)^2 
			9(\ln\alpha+\beta)c_i^{-2}
	}{4}\right)
	=
	\frac1{\alpha} e^{-\beta}.
\end{eqnarray*}
The failure probabilities concerning the assertions about $\|\P_T^\bot
Y\|$ are bounded, as before, by Lemma~\ref{lem:pbot1}. Note 
that we need to employ the ``Poissonian'' part of the lemma, i.e.
Eq.~(\ref{eqn:bernstein2}) when $i>2$.
\begin{eqnarray*}
	p_3(1), p_3(2)
	&\leq&
	\exp\left(
		-\frac{18(\ln(\alpha)+\beta) \ln n}{16  }+\ln(2n)
	\right) \\
	&\leq&
	\frac1{\alpha} e^{-\beta}, \\
	p_3(i) 
	&\leq&
	\exp\left(
		-\frac{\ln n 9(\ln(\alpha)+\beta)}{\sqrt 2}+\ln(2n)
	\right) \\
	&\leq&
	\frac1{\alpha}e^{-\beta}.
\end{eqnarray*}
Lastly, $p_1$ can be comfortably bounded by
\begin{eqnarray*}
	p_1 
	&\leq&
	\exp\left(-\frac{t^2 \kappa }{32 \nu }+\ln(4nr)\right) \\
	&\leq&
	e^{-2(\ln\alpha+\beta)(2\ln n + (l-1))+2\ln n+\ln 4}\\
	&\leq& \frac1{\alpha} e^{-\beta}.
\end{eqnarray*}

A first improved estimate may be achieved at this point by setting
$\alpha=2l$.  From a simple application of the union bound we infer that
the total probability of error is smaller than $e^{-\beta}$.  In
total, the process will have accessed fewer than
\begin{eqnarray}
	&&
	18 (\ln(2l)+\beta)\nu \,4 
	\big(
		2 \ln n + \log_2(2n^2\sqrt r) -2
	\big) nr \nonumber \\
	&=&
	O(nr\nu(\beta+\ln\ln n) \ln n) \label{eqn:reasonablyTight}
\end{eqnarray}
expansion coefficients. 

\begin{theorem}
	Let $\rho$ be a rank-$r$ matrix and suppose that
	$\{w_a\}$ is an operator basis fulfilling $\max_a \|w_a\|^2\leq
	\frac{\nu}{n}$.
	Then the solution $\sigma^\star$ to the
	optimization problem (\ref{eqn:optimization}) is unique and equal to
	$\rho$ with probability of failure smaller than $e^{-\beta}$,
	provided that 
	\begin{equation*}
		|\Omega|\geq O(n r \nu (\beta+\ln\ln n)\ln n).
	\end{equation*}
\end{theorem}

Largely for aesthetic reasons, we provide a further refinement which
does away with the $(\ln\ln n)$-term in (\ref{eqn:reasonablyTight}).
Recall its origin.  Let $p(i)\leq p_1(i)+p_2(i)$ be the probability
that at least one of the two assumptions 
\begin{eqnarray}\label{eqn:iassumptions1}
	&&\|(\P_T \R_i\P_T -\Id)X_{i-1}\|_2<c_i \|X_{i-1}\|_2, \\
	&&\|\P_T^\bot \R_i X_{i-1}\| \leq t_i \|X_{i-1}\|_2
	\label{eqn:iassumptions2}
\end{eqnarray}
made about the $i$th batch does not hold. In the argument
above, we employed the union bound which ascertains that the total
probability of failure is bounded above by $l \max_i p(i)$.  To make
this expression a constant, $\max_i p(i)$ must be $O(l^{-1})$.
This, in turn, was achieved by setting $\alpha=O(\ln l)=O(\ln\ln
n)$.

There is an alternative construction for the dual certificate which
turns out to yield a better estimate. Informally, the idea is to draw
$l'>l$ batches, but to include into the golfing scheme only those
batches for which the assumptions (\ref{eqn:iassumptions1}),
(\ref{eqn:iassumptions2}) hold.  We must choose $l'$ large enough
that, with high probability, $l$ of the batches do fulfill the
assumptions. There is hence a further degree of freedom in the choice
of the parameters: decreasing the $\kappa_i$ increases the average
number of batches not meeting the assumptions, which can be
compensated for by increasing $l'$. It will be shown below that this
freedom may be used to improve the bounds.

To give a formal description of the construction, we re-state the
slightly modified definitions of the objects occurring in the golfing
scheme. The most important change is the introduction of a function
$f:[1,l]\to[1,l']$ which enumerates the batches to be
included.  
More precisely, the objects
\begin{eqnarray*}
	\R_i: \sigma &\mapsto& \frac{n^2}{m_i}
	\sum_{j=m_1+\dots+m_{i-1}+1}^{m_1+\dots+m_i} w_{A_j}\,
	(w_{A_j},\sigma) \\
	\\
  X_0 &=& \sign \rho, \qquad X_i = \sign\rho-\P_T Y_i
\end{eqnarray*}
are defined as before, while
\begin{equation*}
	Y_i = \sum_{j=1}^i \R_{f(j)} X_{j-1}
\end{equation*}
now only depends on a subset of batches. The function $f$, in turn, is
defined by setting $f(0)=0$ and $f(i)$ to
\begin{eqnarray*}
	\min && j>f(i-1) \\
	\text{such that} &&
	\|(\P_T \R_j\P_T -\Id)X_{i-1}\|_2<c_i \|X_{i-1}\|_2, \\
	&&\|\P_T^\bot \R_j X_{i-1}\| \leq t_i \|X_{i-1}\|_2.
\end{eqnarray*}

It remains to choose the parameters of the golfing scheme.  With
foresight, set $\alpha=6$.
Then the probability $p(i)$ of the $i$th batch ($i>2$) being discarded
(i.e.\ $i$ not being in the range of $f$) is smaller than 
\begin{equation*}
	p(i)
	\leq 
	p_1(i)+p_2(i)
	\leq
	\frac1{3}e^{-\beta}. 
\end{equation*}
By the standard Chernoff-Hoefding bound\footnote{E.g.\ Theorem~2.3a in
\cite{mcdiarmid_concentration_1998}; one could also use the Bernstein inequality derived in
this paper, obtaining slightly worse constants.}:
\begin{eqnarray*}
	p_5&:=&
	\Pr
	\left[
	\big(
	\text{number of batches in the range of $f$
	} 
	\big)
	< l
	\right] \\
	&\leq&
	\exp\left(
		\frac{-2(\frac{2}{3}l'-l)^2}{l'}
	\right).
\end{eqnarray*}
We consider this bound in two regimes. First assume that $n\geq
2^{5(\beta+\ln 6)}$ so that $l \geq 2\log_2 n\geq 9(\beta+\ln 6)$.
Choose $l'=2l$.
The exponent becomes $-\frac l9 \leq -(\beta+\ln 6)$.
Next, drop the assumption on $n$ and instead demand $\beta\geq 8+3\ln
6$. Set $l'=\beta\frac 32 l$. In this case a few simple manipulations
yield for the exponent
\begin{equation*}
	-\frac{4(\beta-1)^2 l}{3\beta}\leq -(\beta+\ln 6).
\end{equation*}
In either case:
\begin{equation*}
	p_5 \leq \frac1\alpha e^{-\beta}.
\end{equation*}
By the union bound, the total probability of failure is smaller than
\begin{equation*}
	p\leq p_1+p_2(1)+p_2(2)+p_3(1)+p_3(2)+p_5\leq
	\frac{6}{\alpha}e^{-\beta} = e^{-\beta}.
\end{equation*}

Under the first assumption ($n\geq 2^{5(\beta+\ln 6)}$) the scheme
required knowledge of fewer then
\begin{eqnarray*}
	&&
	18(\ln5+\beta)\nu 4(2\ln n+2\log_2(2n^2\sqrt r))nr\\
	&=&
	O(nr\nu (\beta+1)\ln n)
\end{eqnarray*}
coefficients.  In the second case ($\beta\geq 8+3\ln
6$) the number is
\begin{eqnarray*}
	&&
	18(\ln5+\beta)\nu 4(2\ln n+\frac32\beta\log_2(2n^2\sqrt r))nr\\
	&=&
	O(nr\nu (\beta+1)^2\ln n)
\end{eqnarray*}
Theorem~\ref{thm:tight} follows.

\emph{Remark:} All the arguments of this section remain valid when the
bound on the operator norm of the basis is dropped. The sole
obstruction preventing us from stating $O(r n \nu \ln n)$ bounds for
the more general case is the union bound in
Lemma~\ref{lem:stayIncoherent}. While it seems plausible that one can
overcome this difficulty with reasonable effort, the author has so far
failed to do so.

\subsection{A lower bound}
\label{sec:lower}

Reference \cite{candes_power_2009} gave lower bounds of order $O(n r
\nu \ln n)$ for the number $|\Omega|$ of matrix elements necessary to
fix a rank-$r$ matrix.  Since the theory of low-rank matrix recovery
seems better-behaved for Fourier-type bases, it might be conjectured
that fewer coefficients are sufficient in this case. This hope turns
out not to be realized.

The results of this section imply that the bound of
Theorem~\ref{thm:tight} is tight up to multiplicative constants.

\begin{theorem}\label{thm:stabilizer}
	Let $n=2^k$ be a power of two. Let $\{w_a^{(k)}\}$ be the Pauli
	basis defined in Section~\ref{sec:unitaryBases}.
	\begin{enumerate}
		\item
		Let $\Omega$ be any subset of $[1,n^2]$. If $|\Omega|<(n-2)
		\log_2 n$, then there are two rank-one projections $P_1, P_2$ with
		orthogonal range such that $(w_a, P_1)=(w_a, P_2)$ for all
		$a\in\Omega$.

		\item
		There is a rank-one projection $P_1$ with the following property.
		Let $\Omega$ be a set of numbers in $[1,n^2]$, obtained by
		sampling 
		\begin{equation*}
			m\leq\frac{1}{(1+\epsilon)} n \log_2 n
		\end{equation*}
		times with replacement.  Then with probability
		\begin{equation*}
			p_f \geq
			\left(1-n^{-\frac{\epsilon^2}{2\ln 2\,(1+\epsilon/3)}}\right)
		\end{equation*}
		there exists a rank-one projection $P_2$, orthogonal to $P_1$,
		such that $(w_a, P_1)=(w_a, P_2)$ for all $a\in\Omega$.
	\end{enumerate}
\end{theorem}

The proof makes use of the theory of \emph{stabilizer states}, a
common notion in quantum information theory
\cite{nielsen_quantum_2000}. To make the presentation self-contained,
we have included the briefest outline of this theory as
Appendix~\ref{sec:stabilizer}. The proof below assumes familiarity with
the notions introduced in the appendix.

\begin{IEEEproof}
	In the statement of the theorem, we used a ``one-dimensional'' labeling of
	the Pauli basis elements $w_a$ by numbers $a\in[1,n^2]$. In
	Section~\ref{sec:stabilizer} on stabilizer theory, a
	``two-dimensional'' labeling in terms of pairs $(p,q)$ from
	$[1,n]\times[1,n]$ proved more convenient. We assume that some
	mapping identifying the one set with the other has been chosen and
	will subsequently not distinguish between them.

	For the first statement:

	By Prop.~\ref{prop:mubs}, there are $n$ stabilizer groups $G_x$,
	$x\in\FF_{2^k}$ whose pairwise intersections equal $\{\Id\}$. If
	$|\Omega|$ is smaller than $(n-2)\log_2 n$, then at least one of these
	stabilizer groups intersects $\{w_a\,|\,a\in\Omega\}$ in
	$l<\log_2n=k$ elements.  Call that stabilizer group $G$. By
	Prop.~\ref{prop:stabilizercharacter}, there are distinct
	characters $\chi_1,\chi_2$ of $G$ which agree on $G\cap\{w_a \,|\,
	a\in \Omega\}$. By Prop.~\ref{prop:stabilizers}, $P_{0/1} =
	P(G,\chi_{0/1})$ are two rank-one projectors with orthogonal range.
	By Eq.~(\ref{eqn:projector}), $(w_a, P_1)=(w_a, P_2)$ for
	$a\in\Omega$.

	We turn to the second claim. Take $P_1=P(G_x,\chi)$ for some
	stabilizer group $G_x$ as in Prop.~\ref{prop:mubs} and some
	character $\chi$. As $|G|=n$, the
	probability of a randomly chosen element of the basis to be
	contained in $G$ equals $1/n$. As argued before, there will be an
	orthogonal stabilizer projector $P_2$ compatible with the coefficients
	in $\Omega$, as soon as the intersection between $\Omega$ and
	$\{w_a\,|\,a\in\Omega\}$ is smaller than $k=\log_2 n$. 
	Thus the probability 
	that (\ref{eqn:optimization}) has a unique solution	
	is not
	larger than the probability of an event with probability $1/n$
	occurring at least $\log_2 n$ times in $m= n\log_2 n/(1-\epsilon)$
	trials. This quantity can be bounded by the standard
	Chernoff-Hoefding inequality
	(e.g.~\cite{mcdiarmid_concentration_1998}, Theorem~2.3. (b)). The
	advertised bound follows.
\end{IEEEproof}

\subsection{Non-Hermitian setting}
\label{sec:nonHermitian}

We presented the argument in terms of Hermitian matrices because this
is the natural setting for the Operator-Bernstein inequality. It is,
however, straight-forward to extend the results to arbitrary complex
matrices. The construction in this section serves as a simple proof of
principle; a more refined analysis is certainly possible. 

Indeed, assume both $\rho$ and the $\{w_a\}$ are arbitrary complex
$n\times n$ matrices (in this section, we break with our previous
convention that any matrix is automatically assumed to be Hermitian
unless stated otherwise). We will employ a standard construction
\cite{bhatia_matrix_1997}, associating with any complex $n \times n$-matrix
$\sigma$ a Hermitian $2n\times 2n$-matrix 
\begin{equation}\label{eqn:tilde}
	\tilde \sigma =
	\frac1{\sqrt 2}
	\left(
		\begin{array}{ll}
			0 & \sigma \\
			\sigma^\dagger & 0
		\end{array}
	\right).
\end{equation}
The obvious strategy pursued below consists of the following steps:
\begin{enumerate}
	\renewcommand{\labelenumi}{(\roman{enumi})}
	\item
	from $\{w_a\}$,
	build a suitable Hermitian basis in the space $\mathcal{M}_{2n}$
	of $2n\times 2n$ matrices, 
	\item 
	formulate a matrix recovery problem in terms of 
	$\tilde\rho$ and the basis constructed before, 
	\item 
	compute the incoherence properties 
	of $\tilde\rho$ with respect to
	that basis, 
	\item
	apply the methods detailed in this paper in the extended space, and
	\item 
	show that the original matrix recovery algorithm (i.e.\ the program
	(\ref{eqn:optimization}) applied to $\rho$, $\{w_a\}$ is no more likely to
	fail than the one in the extended space.
\end{enumerate}

To this end, we start by collecting some basic properties of the
mapping $\sigma\mapsto\tilde\sigma$. 

\begin{lemma}\label{lem:tilde}\  \ 
	\begin{enumerate}
		\item
		For $\sigma_1, \sigma_2\in\mathcal{M}_n$:
		\begin{eqnarray}\label{eqn:innerproduct}
			(\tilde \sigma_1, \tilde\sigma_2)
			=\operatorname{Re}\big( (\sigma_1, \sigma_2 \big)\big).
		\end{eqnarray}

		\item
		Let $\{w_a\}_a$ be an ortho-normal basis in the complex vector
		space $\mathcal{M}_n$. Then
		\begin{equation*}
			\{\widetilde w_a \}_a \cup
			\{\widetilde{\mathrm{i}\,w_a}\}_a
		\end{equation*}
		is an ortho-normal basis in the real vector space of Hermitian
		off-diagonal matrices of the form (\ref{eqn:tilde}).

		\item
		Let
		\begin{equation*}
			\sigma = \sum_{i=1}^r s_i\, \psi_i\phi_i^*
		\end{equation*}
		be the singular value decomposition of $\sigma\in\mathcal{M}_n$.
		The $2r$ vectors in $\RR^n\oplus\RR^n$ of the form
		\begin{equation}\label{eqn:eigenvectors}
			\frac1{\sqrt2}\big(\psi_i \oplus  \phi_i\big), \qquad
			\frac1{\sqrt2}\big(\psi_i \oplus  (-\phi_i)\big)
		\end{equation}
		are the normalized non-zero eigenvectors of $\tilde\sigma$, with
		eigenvalues $\frac{\pm 1}{\sqrt 2} s_i$. In particular,
		\begin{equation*}
			\|\tilde\sigma\|=\frac1{\sqrt 2}\|\sigma\|,\quad 
			\|\tilde\sigma\|_1 = \sqrt 2 \|\sigma\|_1,
		\end{equation*}
		and $\rank\tilde\sigma=2\rank\sigma$.

		\item
		With $\sigma$ as above, set
		\begin{equation*}
			E(\sigma)=\sum_{i=1}^r \psi_i \phi_i^*
		\end{equation*}
		(the non-Hermitian analogue of $\sign\sigma$; c.f.\
		\cite{candes_power_2009}).
		Then
		\begin{equation*}
			\sign\tilde\sigma =  \sqrt 2\,\tilde E(\sigma).
		\end{equation*}
	\end{enumerate}
\end{lemma}

\begin{IEEEproof}
	Compute:
	\begin{eqnarray*}
		(\tilde \sigma_1, \tilde\sigma_2)&=&
		\frac12
		\tr
		\left(
			\begin{array}{ll}
				0 & \sigma_1 \\
				\sigma_1^\dagger & 0
			\end{array}
		\right)\,
		\left(
			\begin{array}{ll}
				0 & \sigma_2 \\
				\sigma_2^\dagger & 0
			\end{array}
		\right)\\
		&=&
		\frac12\big( 
		\tr\sigma_1\sigma_2^\dagger
		+
		\tr\sigma_1^\dagger \sigma_2
		\big)
		=\operatorname{Re}\big( (\sigma_1, \sigma_2 \big)\big),
	\end{eqnarray*}
	which implies the first two claims. Verifying statement~3 is trivial.

	Let $\psi_i^{(1)} = \psi_i \oplus 0$, $\phi_i^{(2)}= 0 \oplus
	\phi_i$. Let $P_+$ be the projection onto the positive part of
	$\tilde\sigma$, let $P_-$ project onto the negative part. 
	From (\ref{eqn:eigenvectors}) it follows that $P_\pm$ equals
	\begin{equation*}
		\frac12 \sum_i 
		\psi_i^{(1)}\left(\psi_i^{(1)}\right)^* \pm
		\psi_i^{(1)}\left(\phi_i^{(2)}\right)^* \pm
		\phi_i^{(2)}\left(\psi_i^{(1)}\right)^* +
		\phi_i^{(2)}\left(\phi_i^{(2)}\right)^* .
	\end{equation*}
	Thus
	\begin{equation*}
		\sign\tilde\sigma=P_+-P_-=
		\sum_i 
		\psi_i^{(1)}\left(\phi_i^{(2)}\right)^* +
		\phi_i^{(2)}\left(\psi_i^{(1)}\right)^* 
		=\sqrt 2 \tilde E.
	\end{equation*}
\end{IEEEproof}

We now tackle the first task listed above: building a suitable basis
in $\mathcal{M}_{2n}$. Denote the original basis $\{w_a\}_{a=1}^{n^2}$
by $\mathcal{B}$. The basis $\mathcal{\tilde B}$ in the extended space
is taken to be the set of matrices
\begin{eqnarray*}
	\frac1{\sqrt 2}
	\left(
		\begin{array}{ll}
			0 & w_a \\
			w_a^\dagger & 0
		\end{array}
	\right),
	&\qquad&
	\frac1{\sqrt 2}
	\left(
		\begin{array}{ll}
			0 & {\rm i}\,w_a \\
			-{\rm i}\,w_a^\dagger & 0
		\end{array}
	\right), \\
	\frac1{\sqrt 2}
	\left(
		\begin{array}{ll}
			w_a & 0 \\
			0 & w_a^\dagger
		\end{array}
	\right),
	&\qquad&
	\frac1{\sqrt 2}
	\left(
		\begin{array}{ll}
			{\rm i}\, w_a & 0 \\
			0 & -{\rm i}\, w_a^\dagger
		\end{array}
	\right)
\end{eqnarray*}
for $a=1,\dots, n$. Note that the first two matrices are just
$\widetilde{w_a}$ and $\widetilde{{\mathrm i}w_a}$, so that
Lemma~\ref{lem:tilde}.2 is applicable.

Let $\tilde\Omega$ be a set of $m$ randomly chosen elements from
$\mathcal{\tilde B}$. Below, we will analyze the problem:
\begin{eqnarray}\label{eqn:convexNH}
	\min && \|\sigma\|_1 \label{eqn:optimizationExtended} \\
	\text{subject to} && (\sigma,b_a) = (\tilde\rho,b_a), \quad
	\forall\,b_a \in \tilde\Omega,
	\nonumber
\end{eqnarray}
where the minimization is over all Hermitian matrices $\sigma$ in
$\mathcal{M}_{2n}$. This is step (ii) above. (Note again that we are
interested in the program (\ref{eqn:convexNH}) only as a \emph{means} of
proving that the original program (\ref{eqn:optimization}) works
directly for non-Hermitian objects).

To handle step (iii), 
we introduce further notations.  Let $U=\range\rho,
V=\range\rho^\dagger$ be the row and column space of $\rho$
respectively. Generalizing our earlier definition to non-Hermitian
operators (and following \cite{candes_exact_2009}), let $T$ be the space of
matrices with row space contained in $U$ or column space contained  in
$V$. The projection operator $\P_T$ onto $T$ acts as
\begin{equation*}
	\P_T \sigma = P_U \sigma + \sigma P_V - P_U \sigma P_V.
\end{equation*}
By $\tilde T$ we mean the set of Hermitian matrices in
$\mathcal{M}_{2n}$ with row or column space equal to
$\tilde U=\range\tilde\rho$. Using these notions, the following lemma
relates the incoherence properties of the extended setup to the
original objects.

\begin{lemma}\label{lem:relate}
	\begin{eqnarray}
		\max_{b_a\in\mathcal{\tilde B}} \|b_a\|^2 &=& \frac12
		\max_{w_a\in\mathcal{B}} \|w_a\|^2, \\
		\max_{b_a\in\mathcal{\tilde B}} |(b_a, \sign\tilde\rho)|^2
		&\leq& 2 
		\max_{w_a\in\mathcal{B}} |\big(w_a, E(\rho) \big)|^2, \\
		\max_{b_a\in\mathcal{\tilde B}} \|\P_{\tilde T} b_a\|_2^2
		&\leq&
		\max_{w_a\in\mathcal{B}} \|\P_T w_a\|_2^2.
	\end{eqnarray}
\end{lemma}

\begin{IEEEproof}
	The first two claims follow from Lemma~\ref{lem:tilde}. We prove the
	last statement for $b_a=\tilde w_a$; the other cases are shown
	analogously. We borrow the notation from the proof of
	Lemma~\ref{lem:tilde}.

	Let $P_U^{(1)}$ project onto the span of the $\psi^{(1)}_i$, 
	and $P_V^{(2)}$ onto the span of the $\phi^{(2)}_i$. From
	Lemma~\ref{lem:tilde}, $P_{\tilde U} = P_U^{(1)}+P_V^{(2)}$. Let
	$P^{(1)}$ be the projector onto the first direct summand in
	$\CC^{2n}=\CC^n\oplus \CC^n$, and $P^{(2)}$ onto the second. 
	Then
	\begin{eqnarray*}
		&&\P_{\tilde T}(P^{(1)} \tilde w_a P^{(2)}) \\
		&=&
		P_U^{(1)} \tilde w_a P^{(2)} +
		P^{(1)} \tilde w_a P_V^{(2)} -
		P_U^{(1)} \tilde w_a P_V^{(2)}.
	\end{eqnarray*}	
	From the analogous relation for the adjoint we conclude that
	\begin{equation*}
		\P_{\tilde T} \tilde w_a = \widetilde{\P_T w_a},
	\end{equation*}
	so that the claim follows from Lemma~\ref{lem:tilde}.1.
\end{IEEEproof}

We proceed to steps (iv) and (v). 

\begin{definition}[Coherence, non-Hermitian case]\label{def:coherenceNH}
	The $n\times n$-matrix $\rho$ has \emph{coherence} $\nu$ with
	respect to a basis $\{w_a\}$ if either
	\begin{eqnarray}
		\max_a \|w_a\|^2 &\leq& 2 \nu \frac {1}{n}
	\end{eqnarray}
	or the two estimates
	\begin{eqnarray}
		\max_a \|\P_T w_a\|_2^2 &\leq& 2 \nu \frac{r}{n},  \\
		\max_a \,|\big(w_a, E(\rho)\big)|^2 &\leq& \frac12 \nu \frac{r}{n^2}
	\end{eqnarray}
	hold.
\end{definition}

\begin{corollary}
	The bounds of Theorem~\ref{thm:main} and Theorem~\ref{thm:tight}
	continue to hold for non-Hermitian $\rho$ and $\{w_a\}$, if the
	coherence $\nu$ is measured according to
	Definition~\ref{def:coherenceNH}, and $n$, $r$ are substituted by
	$2n$, $2r$ respectively.
\end{corollary}

\begin{IEEEproof}
	The fact that the problem (\ref{eqn:convexNH}) will have
	$\sigma^\star=\tilde\rho$ as its unique solution with the
	probability of success advertised in Theorems~\ref{thm:main},
	\ref{thm:tight} is an immediate consequence of
	Lemma~\ref{lem:relate}. 

	From Lemma~\ref{lem:tilde}.3, $\|\tilde\rho\|_1\propto \|\rho\|_1$,
	so that the $n\times n$ minimization problem
	(\ref{eqn:optimization}) has $\rho$ as its unique solution whenever
	the same is true for (\ref{eqn:convexNH}) and $\tilde\rho$.
\end{IEEEproof}

\section{Conclusion and Outlook}

\subsection{Outlook}

The following topics will be treated in follow-up publications.

\subsubsection{Noise resilience}

As indicated in \cite{gross_quantum_2009}, the procedures laid out in this paper are
resilient against noise. The analysis of noise effects in the general
case builds on techniques proved in \cite{candes_matrix_2009} for the matrix
completion problem. It turns out that the bounds are quite sensitive
to the operator norm of the sampling operator $\R$ (c.f.\
Eq.~(\ref{eqn:Rnorm})). This number is equal to one if the expansion
coefficients were sampled without replacing, and is likely to be of
order $O(\ln n)\gg 1$ for the i.i.d.\ scheme presented here.  In a
future publication, we will prove operator-valued large deviation
bounds for sampling without replacing \cite{gross__2009}. Therefore, a
detailed discussion of noise effects will be deferred until then.

\subsubsection{Tight frames}

Let $\mu$ be a normalized measure on the unit-sphere of matrices.
We refer to $\mu$ as a \emph{tight frame} (also a \emph{spherical
1-design} or a set of matrices in \emph{isotropic position}
\cite{rudelson_random_1999}, or just an ``overcomplete basis'') if
\begin{equation}\label{eqn:frame}
	\EE_\mu\left[ \P_w\right] :=
	\int  \P_w\, \mathrm{d}\mu(w)= \frac1{n^2}\,\id,
\end{equation}
where $\P_w$ is the orthogonal projection onto $w$.
Tight frames can replace ortho-normal bases in many situations.

In the ``Fourier-type'' case -- i.e.\ if there is a uniform bound on
the operator norm $\|w\|$ of the elements  of the frame -- all
statements in this paper may be easily translated from ortho-normal
bases to tight frames. In the absence of such a constraint,
Lemma~\ref{lem:stayIncoherent} may be a source of problems: it
contains a union bound over all elements of the frame and is therefore
sensitive to its size. In particular, it cannot be directly applied to
continuous frames. We believe that this difficulty can be overcome
with medium effort and may present more details elsewhere.

Note that similar conclusions have been drawn before in the case of
commutative compressed sensing \cite{donoho_compressed_2006}.

\section{Acknowledgments}

The author is glad to acknowledge inspiring discussions with, and the
support of, I.~Bjelakovic, S.~Becker, S.~Flammia, M.~Kleinmann,
V.~Nesme. In particular, he would like to thank J.~Eisert and
Y.-K.~Liu for providing many insights which lead to improvements of
the argument. The manuscript benefited considerably from constructive
comments by the anonymous referees.

\section{Appendix A: proof of Theorem \ref{thm:vecBernstein}}
\label{sec:vectorMartingale}

For completeness, we give a short proof of
Theorem~\ref{thm:vecBernstein} (see also
\cite{ledoux_probability_1991} \cite{yurinskii_exponential_1974}). 

\begin{IEEEproof}
	We aim to use Theorem~\ref{thm:martingale} with $n=1$.  To that end,
	let 
	\begin{equation*}
		Z_i = \EE[N\,|\, \Xb_i] - \EE[N]
	\end{equation*}
	be the Doob martingale sequence of $N-\EE[N]$ with respect to the
	$X_i$; let $X_0=1$. As in Theorem~\ref{thm:martingale}, set $D_i=Z_i
	- Z_{i-1}$.

	Let $\hat \Xb_i$ be the set $\{ X_1, \dots, X_{i-1},
	X_{i+1}, \dots,X_m\}$ of all random variables except for the
	$i$th one. Finally, let 
	\begin{equation*}
		\hat S_i= \sum_{j\neq i} X_{j}
	\end{equation*}
	be the sum of all vectors, with the $i$th term omitted. 

	Using the triangle inequality
	\begin{eqnarray}
		|D_i|
		&=& |\EE[N\,|\, \Xb_i]- \EE[N\,|\, \Xb_{i-1}]| \nonumber \\
		&\leq& 
		\sup_{\hat \Xb_i}\left|
			N- \EE[N\,|\, \hat\Xb_{i}] 
		\right| \label{eqn:vec1} \\
		&\leq& 
		\|\hat S_i\|_2 +\|X_i\|_2 - (\|\hat S_i\|_2 - \EE[\|X_{i}\|_2])
		\nonumber\\
		&=& \|X_i\|_2 + \EE[\|X_i\|_2] \label{eqn:vec2}.
	\end{eqnarray}
	Thus
	\begin{equation*}
		|D_i|\leq \max \|X_i\|_2+\EE[\|X_i\|_2] \leq 2 \max\|X_i\|_2=: c_i,
	\end{equation*}
	where the maximum is over all values of $X_i$.
	With (\ref{eqn:vec1}):
	\begin{eqnarray*}
		\EE[D_i^2\,|\Xb_{i-1}] \leq 
		\sup_{\hat \Xb_i}
		\EE[(N-\EE[N|\hat \Xb_i])^2|\hat\Xb_{i}]. 
	\end{eqnarray*}
	But
	\begin{eqnarray*}
		&&
		\EE\big[
			(N- \EE[N\,|\, \hat \Xb_i])^2 
		\,\big|\, \hat\Xb_{i}
		\big] \\
		&=&
		\EE[ N^2\,|\, \hat\Xb_i ]	-
		\EE[ N\,|\, \hat\Xb_i ]^2 \\
		&=&
		\|\hat S_i\|_2^2 + \EE[ \|X_i\|_2^2 ] -
		\EE[ \|\hat S_i + X_i\|_2 \,|\, \hat\Xb_i ]^2 \\
		&\leq&
		\|\hat S_i\|_2^2 + \EE[ \|X_i\|_2^2 ] -
		\|\hat S_i + \EE[X_i]\|^2
		=
		\EE[ \|X_i\|_2^2]=:\sigma_i^2.
	\end{eqnarray*}

	It remains to compute the expectation $\EE[N]\leq \EE[N^2]^{1/2}$.
	The square of the latter quantity is
	\begin{eqnarray*}
		\EE[N^2] 
		&=&
			\sum_{i,j} \EE[\langle X_i, X_j \rangle] 
		= \sum_i \EE[\|X_i\|_2^2] = V.
	\end{eqnarray*}
\end{IEEEproof}

\section{Appendix B: basic theory of stabilizer states}
\label{sec:stabilizer}

The lower bound in Section~\ref{sec:lower} was built around the
concept of ``stabilizer states'', a concept from quantum information theory.
For the convenience of the reader, we give a short outline below.
The presentation is necessarily both very condensed and fairly
technical. A more complete account can be found e.g.\ in
Refs.~\cite{nielsen_quantum_2000,gottesman_stabilizer_1997,gross_hudsons_2006}.

As a first step, we need to identify a certain group structure of the
elements of the Pauli basis introduced in
Section~\ref{sec:unitaryBases}.

Let $\FF_2$ be the finite field of order
two (with elements $\{0,1\}$), and let $\FF_2^k$ be the set of column
vectors with $k$ entries from $\FF_2$. We introduce a mapping $w$ from
pairs $(p,q)\in(\FF_2^k,\FF_2^k)$ to unitary matrices on
$(\CC^2)^{\otimes k}$ by setting
\begin{equation}\label{eqn:weyls}
	w(p,q) =  
	({\mathrm{i}}^{p_1 q_1}\, \sigma_3^{p_1} \sigma_1^{q_1})\otimes \dots\otimes
	({\mathrm{i}}^{p_k q_k}\, \sigma_3^{p_k} \sigma_1^{q_k}).
\end{equation}

\begin{proposition}[Properties of Pauli operators]
	With $w(p,q)$ as defined in (\ref{eqn:weyls}), it holds that
	\begin{enumerate}
		\item
		The $w(p,q)$'s are Hermitian and unitary. It follows that
		\begin{equation}\label{eqn:squaretoone}
			w(p,q)^2 = \Id.
		\end{equation}

		\item
		The Pauli operators form an super-normalized orthogonal basis:
		\begin{equation}\label{eqn:paulibasis}
			\tr w(p,q) w(p',q') = 2^k\,\delta_{p,p'} \delta_{q,q'}.
		\end{equation}

		\item
		For every $p,q,p',q'$, there is a phase $\lambda(p,q,p',q')\in\{\pm
		1, \pm \mathrm{i}\}$ such that
		\begin{equation}\label{eqn:rep}
			w(p,q) w(p',q') =\lambda(p,q,p',q')\,w(p+p', q+q').
		\end{equation}
		(In other words, the map $w$ realizes a \emph{projective
		representation} of the additive group of $\FF_2^k \times
		\FF_2^k$.)

		If $w(p,q)$ and $w(p',q')$ commute, then (\ref{eqn:rep})
		simplifies to
		\begin{equation}\label{eqn:commuting}
			w(p,q) w(p',q') = \pm w(p+p', q+q').
		\end{equation}

		\item
		The commutation relation
		\begin{equation}\label{eqn:commutation}
			w(p,q) w(p',q') = w(p',q') w(p,q) (-1)^{pq'-qp'}
		\end{equation}
		holds. 
	\end{enumerate}
\end{proposition}

\begin{IEEEproof}
	Equations~(\ref{eqn:squaretoone}, \ref{eqn:paulibasis},
	\ref{eqn:rep}, \ref{eqn:commutation}) can be checked by simple
	direct computation.

	To verify Eq.~(\ref{eqn:commuting}), note that the product of two
	commuting Hermitian operators is Hermitian. Thus, if $w(p,q)$ and
	$w(p',q')$ commute, the l.h.s.\ of
	(\ref{eqn:rep}) is Hermitian. But the r.h.s.\ is Hermitian only if
	$\lambda(p,q,p',q')$ is real.
\end{IEEEproof}

By Eq.~(\ref{eqn:rep}), the set
\begin{equation}
	\P^{(k)}  = \{ \pm w(p,q), \pm \mathrm{i}w(p,q) \,|\, (p,q) \in
	\FF_2^k\times \FF_2^k \}
\end{equation}
forms a matrix group which is known as the \emph{Pauli group}.

Certain subgroups of the Pauli group can be used to define an
interesting class of projection operators. These are called
\emph{stabilizer groups} and defined as follows:

\begin{definition}\label{def:stabilizer}
	Let $G$ be a subgroup of $\P^{(k)}$. The group $G$ is called a
	\emph{stabilizer group} if 
	\begin{enumerate}
		\item
		it is Abelian,
		\item
		$-\Id\not\in G$, and
		\item
		its order $|G|$ equals $2^k$. 
	\end{enumerate}
\end{definition}

The connection between stabilizer groups and projection operators is
given in the next proposition.

\begin{proposition}\label{prop:stabilizers}
	Let $G$ be a stabilizer group. Let $\chi$ be a complex character of
	$G$ (i.e. $\chi(gg') = \chi(g) \chi(g')$ for $g\in G$).  Set
	\begin{equation}\label{eqn:projector}
		P(G, \chi) = \frac1{2^k} \sum_{g\in G} \chi(g) g
	\end{equation}
	Then
	\begin{eqnarray}
		\tr P(G,\chi) &=& 1 \label{eqn:stabilizer:trace} \\
		P(G,\chi)^2 &=& P(G,\chi) \label{eqn:stabilizer:square} \\
		P(G,\chi)^\dagger &=& P(G,\chi)  \label{eqn:stabilizer:hermitian}
	\end{eqnarray}
	In particular, $P(G,\chi)$ is a rank-one projector. 
	
	If $\chi'$ is
	another complex character of $G$, then
	\begin{equation}\label{eqn:stabilizer:orthogonal}
		\tr\Big(P(G,\chi) P(G, \chi')\Big)
		= \delta_{\chi,\chi'}.
	\end{equation}
\end{proposition}

\begin{IEEEproof}
	Equation (\ref{eqn:stabilizer:trace}) follows from
	(\ref{eqn:paulibasis}). 

	Next,
	\begin{eqnarray*}
		P(G,\chi)^2 
		&=& \left(\frac1{2^{k}}\right)^2 \sum_{g,h\in G} \chi(hg) hg \\
		&=& \left(\frac1{2^{k}}\right)^2 |G| \sum_{g\in G} \chi(g) g = P(G,\chi)
	\end{eqnarray*}
	because for $h\in G$ it holds that $h\,G=G$ (which is true for any
	group).

	From Def.~\ref{def:stabilizer}.2 and Eq.~(\ref{eqn:squaretoone}),
	it follows that $g^2=\Id$ for $g\in G$. Hence $\chi(g)^2=1$ so
	that $\chi(g)=\pm 1$. Thus Eq.~(\ref{eqn:projector}) is a real
	linear combination of Hermitian operators and therefore Hermitian.
	This proves Eq.~(\ref{eqn:stabilizer:hermitian}).

	Lastly,
	\begin{eqnarray*}
		\tr P(G,\chi)  P(G,\chi')
		&=& \left(\frac1{2^{k}}\right)^2 \sum_{g,h\in G} \chi(h)\chi'(g)
		\tr hg \\
		&=& \frac1{2^{k}}  \sum_{g\in G} \chi(g)\chi'(g) =
		\delta_{\chi,\chi'}
	\end{eqnarray*}
	having used  Eq.~(\ref{eqn:paulibasis}) and the standard
	orthogonality relation for characters of finite groups (see e.g.
	\cite[Corollary 2.14]{isaacs_character_1976}.
\end{IEEEproof}

Proposition~\ref{prop:stabilizers} allows us to construct rank-one
projection operators from stabilizer groups. It remains to be shown
that such groups actually exist. The construction below makes use of
the fact that $\FF_2^k$ can be identified with the (unique) finite field
$\FF_{2^k}$ of order $2^k$ in the sense that there exists a
(non-unique) isomorphism from $\FF_2^k$ to $\FF_{2^k}$ which respects
the additive structure. In this way, we can assign a meaning to the
product between elements from $\FF_2^k$.

\begin{proposition}\label{prop:mubs}
	Let $b_1, \dots, b_k$ be a basis of $\FF_2^k$.
	For each $x\in\FF_2^k$, let $G_x$ be the subgroup of $\P^{(k)}$
	generated by $\{w(b_1, x b_1), \dots, w(b_k, x b_k)\}$.
	Then $G_x$ is a stabilizer group. If $x'\neq x$,
	then
	\begin{equation}\label{eqn:intersection}
		G_x \cap G_{x'} = \{\Id\}.
	\end{equation}
\end{proposition}

\begin{IEEEproof}
	Since
	\begin{equation*}
		b_i (x b_j) - (x b_i) b_j = 0
	\end{equation*}
	the generators commute mutually by Eq.~(\ref{eqn:commutation}). Thus, $G_x$ is
	Abelian. 
	
	From Eq.~(\ref{eqn:commuting}) it follows that all matrices
	in $G_x$ are of the form $\pm w(p, x p)$ for $p\in\FF_2^k$. This
	proves Eq.~(\ref{eqn:intersection}).

	Combining Eq.~(\ref{eqn:commuting}) with the fact that the $b_i$'s
	are a basis, it is easy to see that for any
	$p\in\FF_2^k$ either $w(p,x p)$ or $-w(p,xp)\in G_x$, but not both.
	Thus $|G_x|=2^k$ and, since $\Id=w(0,x0)\in G$ it must hold that $-\Id\not\in
	G$. Hence $G$ is a stabilizer group and we are done.
\end{IEEEproof}

We need one final statement:

\begin{proposition}\label{prop:stabilizercharacter}
	Any stabilizer group $G$ is isomorphic to the additive group of
	$\FF_2^k$. Given any $l<k$
	elements $\{g_1, \dots, g_l\}$ of G, it is possible to find two
	distinct characters $\chi_1, \chi_2$ of $G$ which agree on $g_1, \dots,
	g_l$.
\end{proposition}

\begin{IEEEproof}
	The first claim follows from (\ref{eqn:commuting}). For the second
	point, note that $g_1, \dots, g_l$ span a subspace of $\FF_2^k$ of
	dimension at most $l<k$. Recall that the complex characters of
	$\FF_2^k$ are in one-one correspondence with linear functionals
	$\FF_2^k \to \FF_2$. There are $2^{(k-l)}$ distinct ways of extending
	a given functional from an $l$-dimensional subspace to all of
	$\FF_2^k$.
\end{IEEEproof}

\bibliographystyle{IEEEtran}
\bibliography{compressed}

\begin{thebibliography}{10}
\providecommand{\url}[1]{#1}
\csname url@samestyle\endcsname
\providecommand{\newblock}{\relax}
\providecommand{\bibinfo}[2]{#2}
\providecommand{\BIBentrySTDinterwordspacing}{\spaceskip=0pt\relax}
\providecommand{\BIBentryALTinterwordstretchfactor}{4}
\providecommand{\BIBentryALTinterwordspacing}{\spaceskip=\fontdimen2\font plus
\BIBentryALTinterwordstretchfactor\fontdimen3\font minus
  \fontdimen4\font\relax}
\providecommand{\BIBforeignlanguage}[2]{{%
\expandafter\ifx\csname l@#1\endcsname\relax
\typeout{** WARNING: IEEEtran.bst: No hyphenation pattern has been}%
\typeout{** loaded for the language `#1'. Using the pattern for}%
\typeout{** the default language instead.}%
\else
\language=\csname l@#1\endcsname
\fi
#2}}
\providecommand{\BIBdecl}{\relax}
\BIBdecl

\bibitem{recht_guaranteed_2007}
\BIBentryALTinterwordspacing
B.~Recht, M.~Fazel, and P.~A. Parrilo, ``Guaranteed {Minimum-Rank} solutions of
  linear matrix equations via nuclear norm minimization,'' \emph{preprint},
  Jun. 2007. [Online]. Available: \url{http://arxiv.org/abs/0706.4138}
\BIBentrySTDinterwordspacing

\bibitem{candes_exact_2009}
E.~Candes and B.~Recht, ``Exact matrix completion via convex optimization,''
  \emph{Foundations of Computational Mathematics}, vol.~9, no.~6, pp. 717--772,
  Dec. 2009.

\bibitem{candes_power_2009}
\BIBentryALTinterwordspacing
E.~J. Candes and T.~Tao, ``The power of convex relaxation: {Near-Optimal}
  matrix completion,'' \emph{preprint}, Mar. 2009. [Online]. Available:
  \url{http://arxiv.org/abs/0903.1476}
\BIBentrySTDinterwordspacing

\bibitem{candes_matrix_2009}
E.~J. Candes and Y.~Plan, ``Matrix completion with noise,'' \emph{Proceedings
  of the {IEEE}}, 2009.

\bibitem{singer_uniqueness_2009}
\BIBentryALTinterwordspacing
A.~Singer and M.~Cucuringu, ``Uniqueness of {Low-Rank} matrix completion by
  rigidity theory,'' \emph{preprint}, Feb. 2009. [Online]. Available:
  \url{http://arxiv.org/abs/0902.3846}
\BIBentrySTDinterwordspacing

\bibitem{keshavan_matrix_2009}
\BIBentryALTinterwordspacing
R.~H. Keshavan, A.~Montanari, and S.~Oh, ``Matrix completion from a few
  entries,'' \emph{preprint}, Jan. 2009. [Online]. Available:
  \url{http://arxiv.org/abs/0901.3150}
\BIBentrySTDinterwordspacing

\bibitem{wright_robust_2009}
\BIBentryALTinterwordspacing
J.~Wright, A.~Ganesh, S.~Rao, and Y.~Ma, ``Robust principal component analysis:
  Exact recovery of corrupted {Low-Rank} matrices,'' \emph{preprint}, May 2009.
  [Online]. Available: \url{http://arxiv.org/abs/0905.0233}
\BIBentrySTDinterwordspacing

\bibitem{donoho_compressed_2006}
D.~Donoho, ``Compressed sensing,'' \emph{{IEEE} Transactions on Information
  Theory}, vol.~52, no.~4, pp. 1289--1306, 2006.

\bibitem{candes_robust_2006}
E.~Candes, J.~Romberg, and T.~Tao, ``Robust uncertainty principles: exact
  signal reconstruction from highly incomplete frequency information,''
  \emph{{IEEE} Transactions on Information Theory}, vol.~52, no.~2, pp.
  489--509, 2006.

\bibitem{candes_near-optimal_2006}
E.~Candes and T.~Tao, ``{Near-Optimal} signal recovery from random projections:
  Universal encoding strategies?'' \emph{{IEEE} Transactions on Information
  Theory}, vol.~52, no.~12, pp. 5406--5425, 2006.

\bibitem{gross_quantum_2009}
\BIBentryALTinterwordspacing
D.~Gross, Y.~Liu, S.~T. Flammia, S.~Becker, and J.~Eisert, ``Quantum state
  tomography via compressed sensing,'' \emph{preprint}, Sep. 2009, to appear in
  Physical Review Letters. [Online]. Available:
  \url{http://arxiv.org/abs/0909.3304}
\BIBentrySTDinterwordspacing

\bibitem{bhatia_matrix_1997}
R.~Bhatia, \emph{Matrix analysis}.\hskip 1em plus 0.5em minus 0.4em\relax
  {New-York}: Springer, 1997.

\bibitem{liu__2009}
Y.~Liu, Jun. 2009, unpublished notes.

\bibitem{werner_all_2001}
R.~F. Werner, ``All teleportation and dense coding schemes,'' \emph{Journal of
  Physics A: Mathematical and General}, vol.~34, no.~35, pp. 7081--7094, 2001.

\bibitem{nielsen_quantum_2000}
M.~A. Nielsen and I.~L. Chuang, \emph{Quantum computation and quantum
  information}.\hskip 1em plus 0.5em minus 0.4em\relax Cambridge University
  Press, 2000.

\bibitem{gottesman_stabilizer_1997}
\BIBentryALTinterwordspacing
D.~Gottesman, ``Stabilizer codes and quantum error correction,'' Ph.D.
  dissertation, Caltech, 1997. [Online]. Available:
  \url{http://arxiv.org/abs/quant-ph/9705052}
\BIBentrySTDinterwordspacing

\bibitem{montanaro_quantum_2008}
\BIBentryALTinterwordspacing
A.~Montanaro and T.~J. Osborne, ``Quantum boolean functions,'' \emph{preprint},
  2008. [Online]. Available: \url{http://arxiv.org/abs/0810.2435}
\BIBentrySTDinterwordspacing

\bibitem{gross_hudsons_2006}
D.~Gross, ``Hudson's theorem for finite-dimensional quantum systems,''
  \emph{Journal of Mathematical Physics}, vol.~47, no.~12, p. 122107, Dec.
  2006.

\bibitem{natarajan_sparse_1995}
B.~K. Natarajan, ``Sparse approximate solutions to linear systems,''
  \emph{{SIAM} J. Comput.}, vol.~24, no.~2, pp. 227--234, 1995.

\bibitem{mesbahi_rank_1997}
M.~Mesbahi and G.~Papavassilopoulos, ``On the rank minimization problem over a
  positive semidefinite linear matrix inequality,'' \emph{Automatic Control,
  {IEEE} Transactions on}, vol.~42, no.~2, pp. 239--243, 1997.

\bibitem{fazel_rank_2001}
M.~Fazel, H.~Hindi, and S.~Boyd, ``A rank minimization heuristic with
  application to minimum order system approximation,'' in \emph{American
  Control Conference, 2001. Proceedings of the 2001}, vol.~6, 2001, pp.
  4734--4739 vol.6.

\bibitem{marshall_inequalities_1979}
A.~W. Marshall and I.~Olkin, \emph{Inequalities}.\hskip 1em plus 0.5em minus
  0.4em\relax Academic Press, 1979.

\bibitem{bertsekas_convex_2003}
D.~P. Bertsekas, A.~Nedić, and A.~E. Ozdaglar, \emph{Convex analysis and
  optimization}.\hskip 1em plus 0.5em minus 0.4em\relax Athena Scientific,
  2003.

\bibitem{ahlswede_strong_2002}
R.~Ahlswede and A.~Winter, ``Strong converse for identification via quantum
  channels,'' \emph{{IEEE} Transactions on Information Theory}, vol.~48, no.~3,
  pp. 569--579, 2002.

\bibitem{recht_simpler_2009}
\BIBentryALTinterwordspacing
B.~Recht, ``A simpler approach to matrix completion,'' \emph{preprint}, Oct.
  2009. [Online]. Available: \url{http://arxiv.org/abs/0910.0651}
\BIBentrySTDinterwordspacing

\bibitem{becker__2009}
S.~Becker, S.~T. Flammia, D.~Gross, Y.~Liu, and J.~Eisert, 2009, in
  preparation.

\bibitem{gross__2009}
D.~Gross and V.~Nesme, 2009, in preparation.

\bibitem{rudelson_random_1999}
M.~Rudelson, ``Random vectors in the isotropic position,'' \emph{Journal of
  Functional Analysis}, vol. 164, no.~1, pp. 60--72, 1999.

\bibitem{petz_survey_1994}
Petz, ``A survey of certain trace inequalities,'' \emph{Functional Analysis and
  Operator Theory}, vol.~30, p. 287, 1994.

\bibitem{dubhashi_concentration_2009}
D.~P. Dubhashi and A.~Panconesi, \emph{Concentration of Measure for the
  Analysis of Randomized Algorithms}.\hskip 1em plus 0.5em minus 0.4em\relax
  Cambridge University Press, Jun. 2009.

\bibitem{tao_additive_2006}
T.~Tao and V.~Vu, \emph{Additive combinatorics}.\hskip 1em plus 0.5em minus
  0.4em\relax Cambridge University Press, 2006.

\bibitem{ledoux_probability_1991}
M.~Ledoux and M.~Talagrand, \emph{Probability in Banach spaces}.\hskip 1em plus
  0.5em minus 0.4em\relax Springer, 1991.

\bibitem{mcdiarmid_concentration_1998}
{McDiarmid}, ``Concentration,'' \emph{Probabilistic methods for algorithmic
  discrete mathematics}, vol.~16, pp. 195--248, 1998.

\bibitem{yurinskii_exponential_1974}
Yurinskii, ``Exponential bounds for large deviations,'' \emph{Theory Probab.
  Appl.}, vol.~19, p. 154, 1974.

\bibitem{christofides_expansion_2007}
D.~Christofides and K.~Markström, ``Expansion properties of random cayley
  graphs and vertex transitive graphs via matrix martingales,'' \emph{Random
  Structures, Algorithms}, vol.~32, no.~1, pp. 88--100, 2007.

\bibitem{tropp_user-friendly_2010}
\BIBentryALTinterwordspacing
J.~A. Tropp, ``User-friendly tail bounds for matrix martingales,''
  \emph{preprint}, Apr. 2010. [Online]. Available:
  \url{http://arxiv.org/abs/1004.4389}
\BIBentrySTDinterwordspacing

\bibitem{isaacs_character_1976}
M.~Isaacs, \emph{Character theory of finite groups}.\hskip 1em plus 0.5em minus
  0.4em\relax New York: Academic Press, 1976.

\end{thebibliography}
\end{document}